\documentclass[conference]{IEEEtran}

\usepackage{amsthm}
\usepackage{amsmath}
\usepackage{amssymb}
\newtheorem{definition}{Definition}
\newtheorem{theorem}{Theorem}
\newtheorem{remark}{Remark}

\usepackage{amsfonts}
\usepackage{caption}
\usepackage{subcaption}
\usepackage{graphicx}
\usepackage[most]{tcolorbox}
\usepackage{xcolor}
\usepackage{algpseudocode}
\usepackage{booktabs}
\usepackage{multirow}
\usepackage[table]{xcolor}
\usepackage[linesnumbered,ruled]{algorithm2e}
\usepackage{makecell}
\usepackage{xurl}
\usepackage{enumitem}

\newcommand{\sssec}[1]{\vspace*{0.08in}\noindent\textbf{#1}}

\usepackage{xcolor}
\newcommand{\hao}[1]{\textcolor{black}{#1}}
\newenvironment{haoblue}
  {\begingroup\color{black}}
  {\endgroup}
\newcommand{\name}{BiasDef}
\newcommand{\attack}{EBI}

\begin{document}

\title{Epistemic Bias Injection:
Manipulating LLM Opinion via Selective Context Retrieval}

\author{
\IEEEauthorblockN{Hao Wu}
\IEEEauthorblockA{\textit{National University of Singapore} \\
hao\_wu@nus.edu.sg}
\and
\IEEEauthorblockN{Prateek Saxena}
\IEEEauthorblockA{\textit{National University of Singapore} \\
dcsprs@nus.edu.sg}
}

\maketitle

\begin{abstract}
When answering user queries, LLMs often retrieve knowledge from external sources stored in retrieval-augmented generation (RAG) databases. These are often populated from unvetted sources, e.g. the open web, and can contain maliciously crafted data. This paper studies attacks that can manipulate the context retrieved by LLMs from such RAG databases. Prior work on such context manipulation primarily injects false or toxic content, which can often be detected by fact-checking or linguistic analysis.
A more subtle threat, which we call epistemic bias injection (\attack), is where adversaries inject factually correct yet epistemically biased passages that systematically favor one side of an open-ended issue. Although linguistically coherent and truthful, such adversarial passages effectively crowd out alternative viewpoints during retrieval from the RAG and push LLM outputs towards an attack-desired stance.

As a core contribution, we propose a novel characterization of the problem: We give a {\em geometric metric that quantifies stance polarity and epistemic bias}. This metric can be computed directly on embeddings of text passages.  Leveraging it, we construct \attack\ attacks and develop a lightweight prototype defense called \name~for them.
We evaluate them both on a comprehensive benchmark constructed from public question answering datasets.
Our results show that: (1) the proposed attack induces significant stance polarity shifts, effectively evading existing retrieval-based sanitization defenses, and (2) \name\ substantially reduces adversarial retrieval and epistemic bias in LLM's answers. Overall, this demonstrates the new threat as well as the ease of employing epistemic bias metrics for filtering in RAG-enabled LLMs.

\end{abstract}

\IEEEpeerreviewmaketitle
\definecolor{darkblueA}{rgb}{0.0, 0.0, 0.4}
\newcommand{\isok}[1]{\ignorespaces#1\unskip}
\newcommand{\tochk}[1]{\textcolor{darkblueA}{\ignorespaces#1\unskip}}

\newcommand{\mysubitem}[2]{\vspace{0.5em}\\\noindent\text{#1)} {\emph{#2}}\ignorespaces}

\definecolor{mygray}{HTML}{e3e6e8}
\definecolor{darkgreen}{RGB}{0, 150, 0}
\definecolor{darkblue}{RGB}{0, 0, 150}
\definecolor{cyanHighlight}{HTML}{5CFFFF}
\definecolor{yellowHighlight}{HTML}{FBFF00}
\definecolor{lightgreen}{RGB}{144,238,144}

\newcommand\prateek[1]{\textcolor{red}{Prateek: #1}}

\section{Introduction}

Retrieval-Augmented Generation (RAG) has become a de facto component of modern large language models (LLMs), enabling systems such as Llama~\cite{huggingface2024llama3}, GPT~\cite{GPT-4}, and DeepSeek~\cite{deepseek} series to incorporate external knowledge at inference time. By retrieving relevant passages from a corpus and adding them into the context, RAG improves performance on open-domain and specialized queries~\cite{healthcare_LLM, chen2021evaluating,deepresearach}.

The use of RAG also introduces an attack surface. RAG databases are often populated from loosely controlled, web-crawled sources rather than curated private knowledge. Injecting likely retrieved content into those sources can manipulate the retrieval results and thereby shape the LLM responses.

Prior work conducted such attacks by injecting malicious passages or prompts into the corpus~\cite{PoisonedRAG,BADRAG,ContextLeakage,illegalToolUse,AgentPoison}, for example via edited Wikipedia pages, deceptive webpages, or news articles~\cite{hanley2024machinemademediamonitoringmobilization}. Such content (e.g., explicit falsehoods, structured prompt injection) within the malicious passages can induce attacker-desired outputs~\cite{PoisonedRAG}, trigger harmful tool usage~\cite{illegalToolUse}, leak sensitive context~\cite{ContextLeakage}, or even cause denial of service~\cite{BADRAG}. Prior defenses~\cite{KAD1,KAD2,WikiCheck,patlan2025realaiagentsfake,LLMRethinking} therefore focus on detecting linguistic fingerprints or eliminating them in the retrieved passages.

\sssec{Opinion Manipulation---A more covert context manipulation}. More recently, Topic-FlipRAG~\cite{TopicFlipRAG} (USENIX Security 2025) highlights a new and subtle class of threats: an adversary who modifies documents in the knowledge database via LLM-guided semantic editing can systematically shift RAG outputs toward a desired opinion. Importantly, Topic-FlipRAG demonstrated that this form of \emph{opinion manipulation} introduces no overt falsehoods and malicious fingerprints. Thus it effectively evades a broad set of existing defenses, including perplexity-based filtering, random masking, paraphrasing, and re-ranking. User experiments further reveal that such manipulation has significant impact on human opinion stance~\cite{TopicFlipRAG}.

\begin{table*}[t]
\centering
\caption{Examples of EBI attacks and their real-world impacts. Each attack injects factually plausible yet epistemically biased passages, shifting the stance polarity of LLM outputs from one position to another---without introducing any detectable falsehoods.}
\label{tab:bias_attack_cases}
\scriptsize
\begin{tabular}{>{\raggedright\arraybackslash}p{2.1cm}|p{4.2cm}|p{5cm}|p{5cm}}
\toprule
\textbf{User Query} & \textbf{Clean LLM Output} & \textbf{Skewed LLM Output (After Manipulation)} & \textbf{Real-World Risk} \\
\midrule
Is the HPV vaccine safe for adolescents?
&
The HPV vaccine has an established safety profile supported by extensive clinical trials and post-market surveillance across millions of doses.
&
While generally considered safe, the HPV vaccine has raised concerns among some experts due to reports of persistent fatigue and potential long-term side effects in adolescents\dots
&
Vaccine hesitancy fueled by LLM misinformation has been linked to measurable declines in HPV vaccination rates, threatening herd immunity against preventable cancers~\cite{kff_health}. \\
\midrule
Should stricter carbon emission regulations be imposed?
&
The scientific consensus supports stricter carbon regulations as a necessary measure to limit global warming; economic modeling suggests net long-term employment gains from green transitions.
&
While beneficial for the environment, stricter regulations may severely impact the economy and lead to widespread job losses in the manufacturing sector, raising questions about their feasibility\dots
&
Shifting public opinion against climate policy weakens political will and electoral support for emissions legislation, directly undermining national and international climate commitments. \\
\midrule
Is cryptocurrency a safe investment?
&
Cryptocurrency markets are highly volatile and speculative; financial regulators worldwide advise caution and recommend it only as a small fraction of a diversified portfolio.
&
Cryptocurrency has proven to be one of the most lucrative investment opportunities in recent years, with early investors in major coins achieving returns far exceeding traditional markets\dots
&
Biased financial advice from LLMs can drive uninformed retail investors toward high-risk positions; crypto market crashes in 2022 wiped out over \$2 trillion in market value, disproportionately harming retail investors \cite{cryptocrash}. \\
\bottomrule
\end{tabular}
\end{table*}

Such opinion manipulation constitutes a security threat~\cite{OWASP_top10, TopicFlipRAG}, especially for AI-augmented search and question-answering systems~\cite{hanley2024machinemademediamonitoringmobilization,xu2024invisiblerelevancebiastextimage,zhou2025exploringescalationsourcebias}. Table~\ref{tab:bias_attack_cases} illustrates concrete attack scenarios. For many open-ended questions in such real LLM applications, there is no single objectively correct or neutral answer---what constitutes a balanced response to ``Should stricter carbon regulations be imposed?'' is itself a matter of perspective. This ambiguity, however, does not diminish the threat; rather, it amplifies it. Precisely because such questions have no ground-truth answer that automated fact-checkers can enforce, an adversary can reliably shift the distribution of LLM responses toward a chosen stance. Such an adversary gains asymmetric influence over how users form their beliefs at a mass scale---a category of risk listed in the OWASP `Top 10 for Gen AI and LLM Risk'~\cite{OWASP_top10}.

\sssec{Unaddressed aspects of existing opinion manipulation}. Topic-FlipRAG demonstrates the viability of this threat but leaves two critical gaps open. First, its attack assumes a privileged adversary with  access to the knowledge database and sufficient domain expertise to specify a target stance and perform elaborate semantic editing---barriers that significantly restrict both the attacker population and the systems at risk. Second, no effective defense is proposed: Topic-FlipRAG's own evaluation shows that all tested mitigations fail, leaving the community without actionable guidance on how to protect RAG systems against opinion manipulation.

\begin{figure*}[t]
    \centering
    \includegraphics[width=0.97\linewidth]{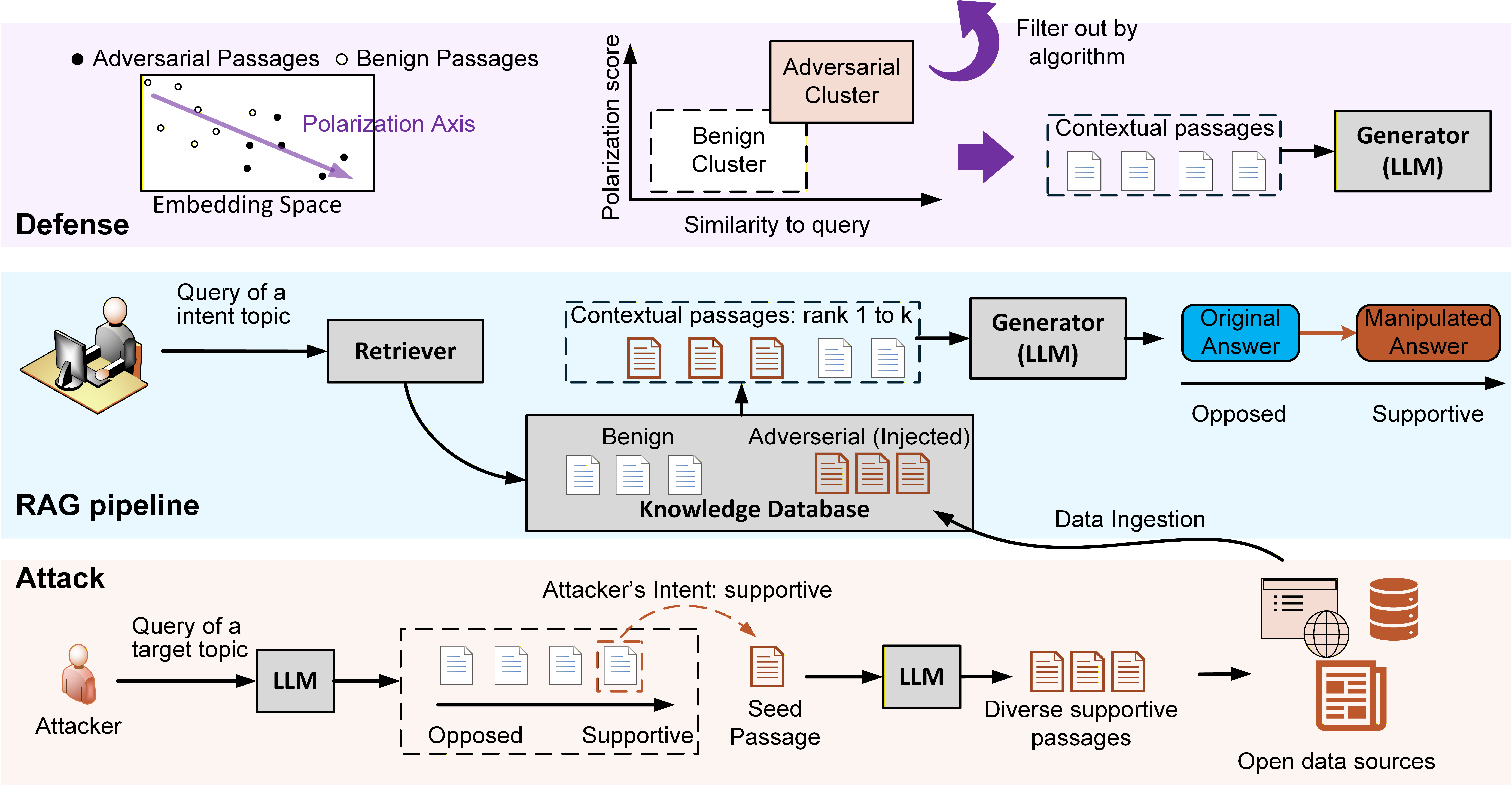}
    \caption{\attack\ attack and \name\ defense in a typical RAG system. A corpus of passages is embedded in a knowledge database. Given a query, the retriever returns the top-$k$ most relevant passages, which are then combined with the query and passed to the generator to produce the final output. The attacker creates adversarial passages with an intended stance and injects them into the corpus by posting them on open data sources. Once ingested by the knowledge database, they influence the contextual passages and the generator's output. \name\ counters this by filtering retrieved passages based on their PS distribution before they reach the generator, reducing epistemic bias in the final response.}
    \label{fig:RAG}
\end{figure*}

\sssec{Our approach}. In this paper, we address both gaps by presenting a geometric view of the problem: the stance polarity (e.g., strongly supporting or weakly opposing) of a passage can be quantified by an explainable geometric metric of its sentence embedding, named \emph{polarization score (PS)}. The PS-shift between two passages reflects the magnitude of their stance polarity difference, enabling us to measure \emph{Epistemic Bias} as the shift in PS before and after the attack. Opinion manipulation can thus be understood as a propagation of PS-shift from retrieved context into the LLM answer---a framing that simultaneously enables a simpler attack and a principled defense. Leveraging this metric, we present (1) \emph{Epistemic Bias Injection (EBI)}, a practical attack that manipulates opinion in any open-domain question without corpus access or domain expertise, and (2) \emph{BiasDef}, a defense that introduces PS directly into the retrieval algorithm (Fig.~\ref{fig:RAG}).

\sssec{EBI attack}. Without access to existing corpus documents or domain expertise, an adversary can still manipulate LLM opinion outputs. The adversary generates passages with diverse viewpoints on a target topic, computes their PS values, and injects those with extreme scores---which represent polar stances---into the corpus. Once retrieved, these passages steer the LLM answer toward the corresponding stance. Crucially, this entire pipeline reduces to a geometric selection criterion on embeddings: no semantic understanding, stance specification, or content-level manipulation is required. The adversarial passages are indistinguishable from naturally authored text, leaving no fingerprint for factual or linguistic defenses to exploit. Our evaluation on various LLMs and open-source datasets~\cite{DUO,hotpotqa,Reddit-dialogues} confirms that existing perspective-aware sanitization defenses~\cite{biasamplify, MMR, SMART-RAG} are largely ineffective against \attack.

\sssec{BiasDef defense}. We also propose \name, a lightweight, plug-and-play defense that augments relevance-based retrieval with PS-based statistical filtering. Because opinion manipulation succeeds by skewing the PS distribution of retrieved passages, the same PS quantification provides a principled defense against this entire class of threats, yielding over a $2.5$--$8.6\times$ reduction in epistemic bias in LLM responses with a formal performance guarantee. \name\ requires no LLM retraining and no labeled data, and is orthogonal to existing model-based defenses. Note that \name\ targets opinion manipulation specifically and is not intended as a universal defense against all forms of RAG poisoning. 

\sssec{Contributions}. Our contributions are as follows:
\begin{itemize}[leftmargin=*, labelsep=0.4em, itemsep=0.2em, topsep=0.2em]
    \item We introduce PS, a continuous and interpretable metric that quantifies stance polarity and epistemic bias in an unsupervised manner, requiring no stance-labeled data and no prior knowledge of the target topic.
    \item We formalize \attack\ attack that operationalizes opinion manipulation under relaxed conditions: no access to existing corpus documents, no domain expertise, and no semantic editing are required. This extends the threat model established by Topic-FlipRAG~\cite{TopicFlipRAG} to a broader and more practical class of adversarial scenarios.
    \item We design \name, a lightweight defense with a formal performance guarantee that leverages PS information to substantially reduce epistemic bias in LLM responses, effective against opinion manipulation.
    \item We construct a RAG benchmark spanning multiple LLMs and public datasets to evaluate both the vulnerability of retrieval algorithms and the effectiveness of \name.
\end{itemize}
\section{Background and Related Work}
\subsection{Primer on RAG}
A typical RAG system is illustrated in Fig.~\ref{fig:RAG}. It comprises a retriever, a generator, and a vector database for knowledge storage. Each passage $d$ in the corpus $D$ is encoded into a fixed-dimensional embedding $E(d)$ using a pre-trained encoder $E(\cdot)$ (e.g., Sentence-BERT~\cite{SBERT}, GTR~\cite{gtr}), and stored in the vector database. When a query $q$ is issued, the retriever identifies the top-$k$ passages with the highest similarity\footnote{Cosine similarity for normalized vectors is directly proportional to their L2 distance, and is commonly used in vector databases~\cite{faiss}.} $\frac{E(q)^TE(d)}{||E(q)||_2\cdot||E(d)||_2}$ to the query embedding $E(q)$.

The retrieved passages are then combined with the query to form a structured prompt for the generator (see Appendix~\ref{app:prompt} for an example). The LLM attends to the retrieved content during decoding, and output quality depends heavily on what is retrieved.

\subsection{Context Manipulation with Fingerprints}
Context manipulation attacks corrupt retrieval results by injecting malicious content into the corpus. Their design involves two sub-problems: (1) ensuring the adversarial passage is retrieved, and (2) crafting content that produces a malicious effect once retrieved.

\sssec{Letting a desired passage be retrieved}. Prior work has proposed various strategies to boost the retrieval ranking of adversarial passages. PoisonedRAG~\cite{PoisonedRAG} and BADRAG~\cite{BADRAG} generate passages highly relevant to a trigger query via LLM prompting and contrastive learning, respectively. AgentPoison~\cite{AgentPoison} crafts an optimal trigger phrase such that any query containing it causes the retriever to prioritize the attacker's documents. While effective at elevating retrieval rank, these approaches lack semantic refinement and can produce unnatural expressions or factual inaccuracies.

\sssec{Crafting malicious content}. Some work~\cite{biasamplify,thota2024attacks,costa2025securingaiagentsinformationflow} focuses on discriminatory bias (e.g., gender or racial) or conceptual bias (e.g., covert brand-name substitution), rather than epistemic bias. Their pipelines may intentionally introduce detectable fingerprints: for example, \cite{biasamplify} directly perturbs passage embeddings, which can introduce semantic errors or factual inconsistencies after decoding. Beyond text injection, prompt injection attacks~\cite{IndirectPrompt,illegalToolUse,DataFlip} insert malicious instructions that override the system's original task. 

Overall, these attacks typically rely on overtly malicious content---factually incorrect statements~\cite{PoisonedRAG,BADRAG}, trigger-bearing spam text~\cite{AgentPoison,BackdoorGramma}, or structured prompts~\cite{IndirectPrompt}---which exposes them to fingerprint-based detection~\cite{KAD1,patlan2025realaiagentsfake,WikiCheck,LLMRethinking}.

\subsection{Opinion Manipulation: No-fingerprint Manipulation}
Topic-FlipRAG~\cite{TopicFlipRAG} proposes a two-stage attack that manipulates the stance of LLM answers by modifying existing documents in the knowledge database. In the first stage, the adversary identifies stance-reflecting information nodes and incorporates them into a target document via iterative lexical substitutions, sentential rewrites, and phrase insertions. In the second stage, a trigger is fused into the modified document to maximize its retrieval relevance across topic queries.

While effective, this pipeline requires the adversary to hold write access to the knowledge database and possess sufficient domain expertise to guide the editing process. Our \attack\ attack removes both requirements, replacing the cumbersome multi-granular editing pipeline with a lightweight, expertise-free selection criterion based on PS.

\subsection{Fingerprint-based Defenses}
Fingerprint-based defenses detect or neutralize adversarial passages by identifying anomalous patterns introduced during attack construction. These methods can be grouped by the signal they exploit.

\sssec{Knowledge and fact consistency}. KAD~\cite{KAD1,KAD2} detects prompt injection attacks by checking whether retrieved passages contain known sentinel answers planted by the defender. Fact-checking defenses~\cite{WikiCheck} validate retrieved passages against trusted references such as Wikipedia. Both are ineffective against opinion manipulation, where adversarial passages carry no injected instructions and are factually correct.

\sssec{LLM-level skepticism}. LLM critical thinking~\cite{LLMRethinking} instructs the generator to critically evaluate retrieved passages before incorporating them into the response, reducing the influence of suspicious content at inference time. However, both Topic-FlipRAG~\cite{TopicFlipRAG} and our results (Sec.~\ref{sec:defense}) demonstrate that this approach fails against semantics-preserving opinion manipulation when no overt falsehood or anomaly is present.

\sssec{Model-based detection}. A separate line of work~\cite{choudhary2025stealthlensrethinkingattacks} identifies adversarial passages by detecting abnormal attention patterns within the LLM. This approach exploits the fact that trigger-bearing attacks concentrate disproportionately high attention weights on specific tokens. Another approach~\cite{patlan2025realaiagentsfake} fine-tunes the LLM on benign reasoning trajectories to suppress memory injection attacks. While effective in their respective settings, these defenses require modifications to the inference framework or model parameters, limiting deployability.

None of the above defenses address opinion manipulation, which introduces no factual errors, syntactic triggers, or anomalous attention patterns. \name\ fills this gap by operating directly on the geometric structure of retrieved passage embeddings.

\subsection{Perspective-based Defense}
Perspective-aware retrievers, such as MMR \cite{MMR}, BRRA \cite{biasamplify}, and SMART \cite{SMART-RAG}, rerank passages by their perspectives rather than semantic relevance alone, as detailed in Appendix~\ref{subsec:existingdefense}. Topic-FlipRAG identifies reranking as the most promising existing mitigation, yet shows it remains insufficient: while these methods reduce the dominance of a single viewpoint in the top-$k$ results, they cannot fully eliminate adversarial content. Our proposed defense \name\ later goes further by directly identifying and filtering adversarial passages via their skewed PS distribution, providing a more targeted defense against opinion manipulation.

\section{Threat Model and Attack Goals}\label{subsec:setup}
\subsection{Quantifying the Stance Polarity of a Passage}\label{subset:metrics}
Human annotation---ranking each passage from strongly opposing to strongly supporting---can quantify the stance polarity of a passage, and thereby measure opinion manipulation as the shift in polarity before and after the attack. However, continuous annotation of large, dynamic corpora is prohibitively expensive and does not scale to real-time retrieval settings. In this paper, we show that a geometric metric can instead be computed in an unsupervised manner from a small set of \emph{anchor passages} with polarizing stances.

Modern LLM sentence encoders map passages into high-dimensional embedding vectors that capture recurring semantic and stylistic patterns. While instance-specific details may be attenuated in the final embedding, latent attributes that are consistent across sentences---such as topic, sentiment, and stance---are preserved. We derive a scalar metric from these high-dimensional embeddings to quantify stance polarity.

\sssec{Key Insight}. Our hypothesis is that the {\em difference in the stance of a given set of text passages, which are otherwise similar (e.g. supporting the same thing), is geometrically discernibly in some direction of the embedding space}. 
Identifying that direction and projecting embeddings onto it yields a scalar value that reflects both the orientation and intensity of a stance, enabling quantitative comparison across passages.

\begin{figure*}
    \centering
    \includegraphics[width=\linewidth]{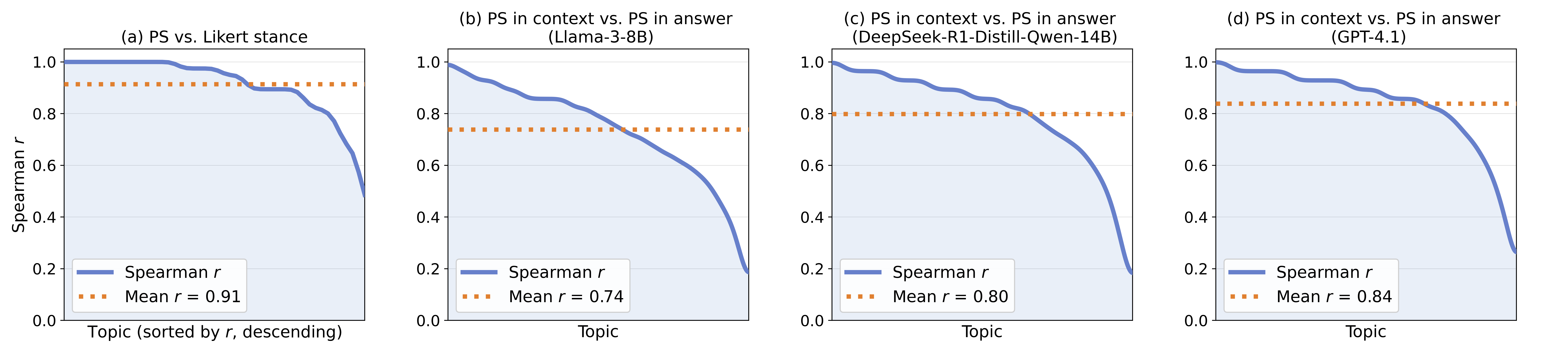}
    \caption{(a) PS values versus 7-point Likert stance scores (annotated by GPT-5.2) for 350 passages drawn from 50 randomly selected questions spanning social, cultural, and technological topics. PS achieves an average Spearman’s rank correlation of 0.91 with Likert scores, demonstrating a strong monotonic alignment with judgments of stance polarity. (b) (c) (d) Correlation of epistemic bias in retrieved passages and the LLM response across  452 topics in WIKI-BALANCE~\cite{DUO}. The high Spearman’s rank correlation shows that epistemic bias can be effectively propagated through different LLMs~\cite{huggingface2024llama3, deepseek2024r1distill,GPT-4} to responses.}
    \label{fig:LLM-as-a-judge}
\end{figure*}

To identify such a direction, we take a set of \emph{anchor passages} with diverse stances on the \emph{same topic}, obtain their embedding vectors via an LLM encoder, and perform principal component analysis (PCA)~\cite{PCA} on them. The leading principal component---which geometrically maximizes the average variance of projections in the Euclidean $\ell_2$ sense---is identified as the \emph{polarization axis}. Because the topic is held fixed while only the stance varies across anchor passages, the dominant source of variance among their embeddings is stance polarity itself, ensuring that this axis captures stance rather than topic relevance or writing style---as evidenced in Fig.~\ref{fig:LLM-as-a-judge}(a) later.


Formally, given $J$ anchor passages $d_{\text{anchor},j}$ associated with a specific question, we compute their embeddings and stack them into a matrix $\mathbf{E}\in\mathbb{R}^{J\times \ell}$, where
$e_j = E(d_{\text{anchor},j}) \in \mathbb{R}^{\ell}$.
The polarization axis $\mathbf{u}_{\text{polar}}$ is obtained by solving
\begin{equation}
    \mathbf{u}_{\text{polar}} =
    \mathop{\arg\max}_{\|\mathbf{u}\|=1}
    \mathrm{Var}(\mathbf{E}\cdot\mathbf{u}),
\end{equation}
which corresponds to the eigenvector associated with the largest eigenvalue of the sample covariance matrix of $\mathbf{E}$. This is the leading principal component, or the polarization axis.

Each new text passage can then be projected onto the polarization axis. For a passage $d$, its embedding projected onto this axis produces a scalar \emph{polarization score} (PS):
\begin{equation}
    \mathrm{PS}(d)= \mathbf{u}_{\text{polar}}^\top E(d),
\end{equation}

A useful property that the PS metric satisifies is \textbf{local consistency.} Consider two passages $d_1$ and $d_2$ about a topic and let $\{\mathbf{u}_k\}$ denote the orthonormal PCA basis. The squared $\ell_2$ distance between embeddings can be decomposed as:
    \begin{equation}
    \begin{aligned}
        \|E(d_1)-E(d_2)\|_2^2
        &= \sum\nolimits_k \bigl((E(d_1)-E(d_2))^\top \mathbf{u}_k\bigr)^2 \\
        &\ge \bigl(\mathrm{PS}(d_1)-\mathrm{PS}(d_2)\bigr)^2.
    \end{aligned}
    \end{equation}
    Therefore, if two passages have high similarity---a small Euclidean distance in their embeddings, their difference in PS score is small. We call this property local consistency.


\sssec{Does PS capture stance polarity?} We validate PS against the LLM-assisted Likert stance score used in Topic-FlipRAG~\cite{TopicFlipRAG}. The reliability of this LLM-as-judge scoring has itself been validated against human annotation in~\cite{TopicFlipRAG}. Specifically, we randomly select 350 passages covering 50 questions from a public dataset~\cite{DUO} and ask GPT-5.2 to rank each passage on how strongly it supports or opposes the question. We then compute PS values for the same passages using the \texttt{sentence-t5-xl}~\cite{t5xl} encoder. Fig.~\ref{fig:LLM-as-a-judge}(a) shows that PS consistently correlates with the LLM ranking across topics, achieving a strong Spearman rank correlation~\cite{Spearman}---suggesting that PS captures the notion of epistemic bias comparably to human-aligned LLM judgment. This evaluation also confirms that passages expressing opposing and supportive stances tend to lie on opposite sides of the mean along the polarization axis. 





\subsection{Threat Model}\label{subsec:threat}
We present the threat of opinion manipulation in our new geometric view in the embedding space. Given a query $q$ and a corpus $\mathcal{D}$, the LLM generates a response based on both the query and a subset of $k$ retrieved contextual passages $R_k(\mathcal{D}|q)\subseteq \mathcal{D}$. We denote the response by $\mathcal{A}(q,R_k(\mathcal{D}|q))$. 

\sssec{Objective of adversary}. The adversary injects a group of adversarial passages $\mathcal{D}_\text{adv}$ into the corpus, forming a poisoned corpus $\mathcal{D}'=\mathcal{D}\cup\mathcal{D}_\text{adv}$, thereby causing desired PS shift into the retrieved passages and the response. We formally define such PS shift as epistemic bias.



\begin{definition}[Epistemic Bias]
\label{def:EBI}
Let $\mathcal{D}'_q = \mathcal{D}_q \cup\mathcal{D}_{\mathrm{adv}}$ 
denote the poisoned corpus, where $\mathcal{D}_{\mathrm{adv}}$ is a 
set of adversarially injected passages. The epistemic bias in retrieved passages and the LLM response is denoted by the shift in their (average) PS values caused by $\mathcal{D}_{\mathrm{adv}}$:
\begin{align}
    &B_\text{retrieval}:=\mathbb{E}_{d\in R_k(\mathcal{D}'|q)}\text{PS}(d)-\mathbb{E}_{d\in R_k(\mathcal{D}|q)}\text{PS}(d),\\
    &B_\text{response}:=\text{PS}(\mathcal{A}(q,R_k(\mathcal{D}'|q)))-\text{PS}(\mathcal{A}(q,R_k(\mathcal{D}|q))).
\end{align}
\end{definition}

\begin{remark}[Agnosticism to normative neutrality]
\label{remark:agnostic}
Definition~\ref{def:EBI} makes no reference to an externally defined ``neutral'' standpoint. An attacker who shifts retrieval outcomes away from this distribution is mounting an EBI attack regardless of whether the shift moves toward or away from any particular stance. This directly addresses the 
concern that ``unbiased'' content may itself be perceived as biased: our framework does not require consensus on what constitutes a neutral position. The threat is the covert, unauthorized redistribution of retrieved stances by an external adversary.
\end{remark}

For any target open-ended topic, the adversary aims to skew the stance polarity of the LLM response by crafting factually correct yet epistemically biased passages. The feasibility of such an attack rests on the following observations.  

\begin{tcolorbox}[colback=green!6, colframe=black, left=1mm, right=1mm, top=1mm, bottom=1mm, boxrule=0.5pt]
\textbf{Observations (Fig.~\ref{fig:LLM-as-a-judge}(b,c,d))}. \emph{Retrieval bias propagates through the generation process. LLM models fail to mitigate the injected epistemic biases during answer synthesis, even when equipped with critical thinking instructions in the prompt (see Prompt-Generation in Appendix~\ref{app:prompt} for the prompt used). Merely scaling up the model or switching architectures does not inherently improve robustness.}
\end{tcolorbox}

\sssec{Capability of adversary}. The LLM encoder, indexed benign passages, and model parameters are fixed and unavailable to the attacker; the adversary cannot modify the retriever, alter similarity metrics, or tamper with pre-existing database entries. The adversary's sole capability is injecting a limited number of new passages into the knowledge corpus---for instance, by publishing content on publicly accessible sources that the RAG pipeline ingests. This is a strictly weaker assumption than Topic-FlipRAG~\cite{TopicFlipRAG}, which requires write access to existing database entries, and better reflects real-world scenarios where the RAG system is operated by a third party.

\section{\attack\ Attack on RAG Databases}
\label{sec:attack_model}
The observations in Sec.~\ref{subsec:threat} confirm that injecting passages into the corpus can propagate epistemic bias through retrieval into the LLM answer. We now present an automated attack that realizes this for arbitrary queries, with the entire pipeline driven by geometric PS computation rather than human expertise.

\subsection{Security Properties}\label{subsec:properties}

An adversarial passage $d_{\text{adv}}$ should have $3$ properties below to work effectively.

\sssec{Property 1---High relevance}.  LLMs have a limited-length input context, especially in locally deployed small-scale settings due to constraints in model capacity, memory consumption, and generation latency. Hence, only a small subset of the retrieved passages from RAG corpus can be included in the input context. So, attackers must ensure their adversarial passages rank highly in their relevance to the query (i.e., cosine similarity) during retrieval. Stated formally, we want: $\text{Sim}(q,d_{\text{adv}}) > \text{Sim}(q, d_{\text{benign}})$, so that $d_{\text{adv}}$ is prioritized during retrieval over other benign passages $d_{\text{benign}}$. 

\sssec{Property 2---Epistemic bias}. The adversarial passage must express a viewpoint that deviates from the baseline distribution of retrieved passages under no attack. To maintain a consistent bias, multiple adversarial passages from a single attack must all have PS values either consistently higher or consistently lower than those of benign passages. This property is expressed as $\text{PS}(d_\text{adv}) > (\text{or} <)\mathbb{E}_{d\in R_k(\mathcal{D}|q)}\text{PS}(d)$.

\sssec{Property 3---Stealth (No explicit fingerprints)}. 
The adversarial passages must avoid introducing explicit or easily detectable fingerprints that distinguish them from benign content.
In particular, they should remain factually correct, linguistically natural, and free of overt signals such as false statements, profanity, or instruction-like prompts.

\subsection{Workflow of \attack\ attack}
The core of \attack\ is to produce human-like yet epistemically biased text for a target question. Because the attacker controls the content fully, we can use an LLM to help generate these passages. Prior work suggests that dense retrievers tend to favor LLM-written content in retrieval~\cite{BiasAndFair, Dai_2024}, so we can expect that using a state-of-the-art LLM to produce text will be effective. The attack process for a target question follows several steps:
\begin{itemize}[leftmargin=*, labelsep=0.4em, itemsep=0.2em, topsep=0.2em]
    \item (1) The attacker first needs $J$ passages with different perspectives on the topic in query $q$, to serve as \emph{candidate passages} for the attack. One approach is to retrieve these passages via keyword search using search engines. In practice, the attacker may prompt an LLM to generate them, using the Prompt-Synthetic in Appendix~\ref{app:prompt} as the prompt: 
    \begin{equation}
        \{d_{\text{synt},j}\}_{j\in[1,J]} = \textbf{LLM}(\text{Prompt-Synthetic}, q),
    \end{equation}
    where $\textbf{LLM}$ represents the LLM generation process. We use the latter.
    
    \item (2) The attacker employs an encoder $E(\cdot)$ to project the candidate passages into the embedding space and performs PCA~\cite{PCA} on the embeddings $\{E(d_{\text{synt},j})\}_{j\in [1,J]}$, without requiring any stance labels. In other words, these candidate passages also serve as the anchor passages to derive the polarization axis. By projecting the passage embeddings onto the resulting polarization axis, the attacker obtains their PS values $\{\text{PS}(d_{\text{synt},j})\}_{j\in [1,J]}$.
    
    \item (3) If the attacker has a preferred stance and the expertise to identify aligned passages, he can select the corresponding $d_{\text{bias}}$ from $\{(d_{\text{synt},j})\}_{j\in [1,J]}$ directly. However, such expertise is not required for an effective attack. An \textbf{expertise-free} attacker wishing to skew the stance polarity of LLM answers need only select the passage $d_\text{bias}$ with the highest epistemic bias (i.e., $\text{PS}(d_{\text{synt},j})-\mathbb{E}_{d_{\text{synt},j}}\text{PS}(d_{\text{synt},j})$) within the candidate passages\footnote{This proxy is valid because Prompt-Synthetic  elicits passages covering diverse stances, so their PS values span both polarity sides and their mean approximates a balanced baseline. Empirically, across all 452 queries in WIKI-BALANCE, the deviation $|\mathbb{E}_{j}\,\mathrm{PS}(d_{\mathrm{synt},j}) - \mathbb{E}_{d \in R_k(\mathcal{D}|q)}\mathrm{PS}(d)|$ averages $\le 0.02$, negligible compared to the attack-induced PS shift in Fig.~\ref{fig:recall}(d).} by choosing the passage with the most extreme PS:
    \begin{equation}
        d_\text{bias} = \mathop{\arg\max}_{d_{\text{synt},j}} \text{PS}(d_{\text{synt},j}) \text{ or } \mathop{\arg\min}_{d_{\text{synt},j}} \text{PS}(d_{\text{synt},j})
    \end{equation}
    inducing a positive or negative polarity shift\footnote{The terms ``positive'' and ``negative'' distinguish opposing directions of stance (e.g., supporting vs.\ opposing a policy) and carry no evaluative or emotional connotation.} in the generated answer. For convenience, we refer to these as \emph{positive \attack} and \emph{negative \attack}, respectively.
    
    \item (4) Using $d_\text{bias}$ as a stance reference, the attacker prompts the LLM to generate $n$ passages on the question $q$ that share the same stance polarity as $d_\text{bias}$ while varying freely in evidence, framing, and language style:
    \begin{equation}
        d_{\text{adv},j} = \textbf{LLM}(\text{Prompt-Adv},q,d_\text{bias}), \quad j\in[1,n].
    \end{equation}
    To ensure factual accuracy and semantic coherence, we carefully design Prompt-Adv (see Appendix~\ref{app:prompt}) to guide generation toward grounded and contextually relevant content. Each generated passage then undergoes a post-generation PS check; only passages whose PS values are sufficiently close to that of $d_\text{bias}$ are retained. This process yields passages that \textbf{share the same stance polarity as $d_\text{bias}$ but are textually distinct from both $d_\text{bias}$ and one another}. Importantly, diversity beyond the primary stance polarity dimension is not only permitted but encouraged, as it better reflects the natural variation of real-world text and hinders retrieval-level deduplication mechanisms that penalize inter-passage similarity (Sec.~\ref{sec:defense}).


    \item (5) Finally, the $n$ adversarial passages are posted on open data sources (e.g., WIKI pages, blogs, and third-party datasets) that can be ingested by RAGs through web search or scraping. Once ingested and retrieved, these passages introduce epistemic bias into the generation process.
\end{itemize}


\sssec{Satisfying security properties in practice}. The above workflow satisfies the three security properties in Sec.~\ref{subsec:properties} by construction: topic-conditioned generation ensures strong 
semantic alignment between adversarial passages and the query 
(Property~1)~\cite{BiasAndFair, Dai_2024}; constraining the stance polarity of the injected passages ensures consistently injection of epistemic bias (Property~2); and explicit instructions for factual correctness 
and linguistic coherence prevent detectable fingerprints 
(Property~3, validated in Sec.~\ref{subsec:securityPro}).



\sssec{Stance polarity versus multi-dimensional viewpoint diversity.}\label{subsec:multidim}
In most open-ended debates, the principal axis of disagreement is inherently binary: a passage either leans toward supporting or opposing a position. Finer-grained dimensions---the choice of evidence, rhetorical framing, or linguistic style---modulate \emph{how} a stance is expressed, not \emph{which} side it takes. PS directly captures this dominant axis, and manipulating it is sufficient to produce a consistent stance polarity shift in LLM answers, as our experiments in Sec.~\ref{sec:defense} confirm. Secondary dimensions such as framing and evidence selection, while contributing to surface diversity, are unlikely to constitute an independent attack axis: varying \emph{how} a stance is argued without shifting the underlying stance polarity is unlikely to produce a systematic change in user belief.

This design also yields an important practical benefit. Because generated passages are constrained only along the primary stance polarity dimension and vary freely in all others, they are textually heterogeneous by construction. This mirrors the natural diversity of real-world web content and defeats retrieval-level deduplication mechanisms---such as MMR~\cite{MMR}---that suppress redundant passages by penalizing high inter-passage similarity (see Sec.~\ref{sec:defense}).

\section{Experiments}\label{sec:defense}

\subsection{Benchmark for Measuring \attack~Attacks}\label{subsec:benchmark}
We built an RAG benchmark to evaluate the effectiveness of existing methods against our proposed \attack\ attack.

\begin{haoblue}
\sssec{Datasets.} Our experiments are mainly based on an open-source dataset, WIKI-BALANCE~\cite{DUO}, containing 452 open-ended, controversial questions across diverse topics and a corpus of 4662 real Wikipedia pages relevant to these questions. While using WIKI-BALANCE as the primary dataset, we also applied our attack and defense methods to Reddit-Dialogues\footnote{Each query in this dataset includes a Reddit topic and previous responses of users, the RAG generates a new response based on retrieved passages from a given knowledge database.}~\cite{Reddit-dialogues} and HotpotQA\footnote{We use ChatGPT-5 to extend some closed-form questions (e.g., "Were Scott Derrickson and Ed Wood of the same nationality?") to open-ended questions ("How did Scott Derrickson’s and Ed Wood’s cultural and national backgrounds shape their filmmaking styles and themes?").}~\cite{hotpotqa} to show the generalization across multiple datasets.
\end{haoblue}

\sssec{RAG setup.} We implement a RAG system on a server equipped with 4$\times$NVIDIA A40 GPUs. The system consists of three components:
\begin{itemize}[leftmargin=*, labelsep=0.4em, itemsep=0.2em, topsep=0.2em]
    \item \emph{Knowledge database}: The database initially indexes benign passages from one of the three public datasets.  
    
    \item \emph{Retriever}: We use \emph{msmarco-distilbert-base-tas-b}~\cite{reimers2020msmarco} as the encoder for our dense retriever. This pre-trained model is fine-tuned on large-scale passage ranking datasets using contrastive learning and has demonstrated strong performance in dense retrieval tasks~\cite{SBERT}. During retrieval, both queries and documents are encoded into dense vectors, and their relevance is computed via cosine similarity. The retriever first coarsely selects the top-$4\times k$ most relevant candidate passages from the knowledge database, and then applies existing methods to filter the final top-$k$ passages.
    
    \item \emph{Generator}: We deploy LLMs with various model sizes—including Meta-Llama-3-8B \cite{huggingface2024llama3}, DeepSeek-R1-Distill-Qwen-14B \cite{deepseek2024r1distill}, and GPT-4.1 \cite{GPT-4} (via the OpenAI API)—using the vLLM framework~\cite{vllm} to generate answers to open-ended questions. A prompt containing the query, the $k$ retrieved passages, and critical thinking instructions is constructed to initiate the answer generation process (see Prompt-Generation in Appendix~\ref{app:prompt}). 
\end{itemize}

\sssec{Attacker settings}. \hao{As outlined in Sec.~\ref{sec:attack_model}, the attacker deploys a DeepSeek-R1-Distill-Qwen-14B model to generate $J=8$ synthetic passages. These passages are then evaluated using a separate \texttt{sentence-t5-xl}~\cite{t5xl} encoder to calculate their PS values, as the RAG's own encoder is unknown to the attacker. We assume an expertise-free attacker who selects the passages with the highest and lowest PS values as the two adversarial passages. Based on the biased passages, the attacker prompts the model to generate $n=10$ adversarial variants for positive \attack\ and another 10 for negative \attack\ and finally injects them into the knowledge database.}

\sssec{Methodology}: 
\hao{In each experimental round, we target one query and evaluate all methods under both positive and negative EBI attacks at three injection numbers: 1, 5, and 10 adversarial passages. These numbers correspond to approximately 0.1–1$\times$ the average number of benign passages per query for WIKI-BALANCE and HotpotQA, and 0.02–0.2$\times$ for Reddit-Dialogues, respectively. We repeat this procedure for all queries in the corpus and report the average performance. For each query, we generate answers three times using the same retrieved context and decoding parameters. This repetition mitigates the influence of occasional outlier responses caused by the stochastic decoding process of LLMs.}

We set $k=5$ as the default number of retrieved passages throughout our evaluation. This choice follows common practice in RAG systems, where a small number of high-quality contexts is preferred to balance informativeness and computational efficiency. Specifically, typical retrieved passages are long articles (e.g, Wikipedia pages), many of which exceed 10K tokens. A larger $k$ would slow down inference, introduce more distractors (i.e., irrelevant content), and may even cause the total context length to exceed the input token limit of our locally deployed models.

\hao{Apart from the results about generalization across LLM models and datasets, we present attack efficacy results only for the most representative setting: DeepSeek-R1-Distill-Qwen-14B~\cite{deepseek2024r1distill} as the generator and WIKI-BALANCE as the evaluation dataset~\cite{DUO}. The complete code and data are released in our artifact.}

\sssec{Evaluation metrics.} We use the following metrics:

\begin{itemize}[leftmargin=*, labelsep=0.4em, itemsep=0.2em, topsep=0.2em]
    \item \textbf{Adversarial Recall (A-Recall)@k}: The proportion of attacker-injected adversarial passages appearing among the top-$k$ retrieved results. This metric evaluates the effectiveness of the attack in getting adversarial content retrieved. A successful attack aims to maximize this rate to crowd out benign content and influence the generated answer toward the attacker’s desired perspective. Conversely, the defense mechanism aims to minimize it.
    
    \item \textbf{Recall@k}: The proportion of correctly identified relevant benign passages among the top-$k$ results, assessed using \textit{Qrels}\footnote{Qrels (manual relevance judgments) are standard evaluation resources that contain query-document pairs labeled with binary or graded relevance scores. These labels are annotated by assessors following strict guidelines and are widely used as ground truth for evaluating retrieval quality.}. A higher Recall@k implies that the retrieved context is more informative and aligned with the intended question.
    
    \item \textbf{Polarization Score (PS)}: As defined in Sec.~\ref{subset:metrics}, PS quantifies the viewpoint of a passage. PS shift---the magnitude of the difference between the PS values in the attacked and unattacked cases---measures the impact of attacks on the bias level of the content. 
    
\end{itemize}

\sssec{Compared baselines}. We evaluate against the perspective-aware retrieval defenses described in Sec.~\ref{subsec:existingdefense}, which represent the most promising class of mitigations identified by Topic-FlipRAG~\cite{TopicFlipRAG}: among all tested defenses, retrieval-stage reranking with perspective awareness showed ability to suppress opinion manipulation, making it the strongest baseline for our evaluation.

\begin{itemize}[leftmargin=*, labelsep=0.4em, itemsep=0.2em, topsep=0.2em]
    \item \textbf{No defense (No Def.)}: Directly retrieves the top-5 passages with the highest cosine similarity to the query~\cite{DR}. 

    \item \textbf{BRRA}~\cite{biasamplify}: BRRA introduces Gaussian noise into the query embedding to generate multiple perturbed variants. we set the noise intensity to 1.0 (i.e., the perturbation magnitude equals the original embedding norm) to maximize variant diversity. Passages relevant to the original and perturbed queries are retrieved, and the union of these results is re-ranked based on retrieval frequency and rank across all queries. The top-5 passages after re-ranking are then selected as the contextual input to the LLM.
    
    \item \textbf{MMR}~\cite{MMR}: We deploy the native MMR algorithm and set $\lambda$---the coefficient that balances the relevance and diversity of the retrieved passage---to a commonly used value: 0.5.

    \item \textbf{SMART}~\cite{SMART-RAG}: As described in \cite{SMART-RAG}, we construct a conflict-aware kernel matrix based on three components: (i) the query-context relevance matrix (measured via cosine similarity between each passage and the query), (ii) the similarity matrix between passages, and (iii) a conflict matrix derived NLI predictions. For conflict detection, we use a pre-trained NLI model\footnote{Available at https://huggingface.co/MoritzLaurer/mDeBERTa-v3-base-xnli-multilingual-nli-2mil7.}. This kernel matrix is then used to the described algorithm for passage ranking.
    \item \textbf{Automated Fact-Checking}~\cite{WikiCheck}: Unlike Q\&A domains where a golden source of facts/truth is available (e.g., medical data~\cite{han2024medical}), we are working with queries that have subjective opinions as answers. As a baseline defense, fact-checking to check for overt falsehoods can be employed.
    \hao{Existing automated fact-checking APIs provide per-claim (i.e., a complete sentence) check yet; thus, they cannot apply to long passages directly. We empirically compare the benign and adversarial passages, broken down into sentences, to test if fact-checking can distinguish between the two.}
\end{itemize}


\subsection{Security Properties of EBI Attacks}\label{subsec:securityPro}
To evaluate how well the crafted adversarial passages satisfy the security properties in Sec.~\ref{subsec:properties}, we compare the similarity-to-query and PS values of adversarial passages to those of benign passages in the three real datasets. In addition, we apply fact checking on the crafted passages to evaluate their stealth.
\begin{figure}[t]
  \centering
  \begin{minipage}[t]{\linewidth}
    \centering
    \captionof{table}{Security property check of adversarial passages generated by our workflow on three datasets.}
    \label{tab:attackratio}
\resizebox{\linewidth}{!}{
    \begin{tabular}{l|l|ccc c ccc}
      \toprule
      Dataset & EBI Type & {Both} & {Only epistemic bias} & {Only high relevance} & {Neither} \\
      \midrule
      \multirow{2}{*}{\shortstack{WIKI-\\BALANCE}} & Negative & 87.5\% & 9.8\% & 2.4\% & 0.3\% \\
      & Positive & 86.0\% & 12.1\% & 1.7\% & 0.2\% \\ 
      \midrule
      \multirow{2}{*}{HotpotQA} & Negative & 75.8\% & 3.6\% & 20.2\% & 0.3\% \\
      & Positive & 71.8\% & 3.7\% & 24.0\% & 0.5\% \\ 
      \midrule
      \multirow{2}{*}{\shortstack{Reddit-\\Dialogues}} & Negative & 50.2\% & 16.0\% & 29.4\% & 4.3\% \\
      & Positive & 43.9\% & 12.4\% & 36.9\% & 6.8\% \\  
      \bottomrule
    \end{tabular}}
  \end{minipage} \\ 
\end{figure}

\begin{figure}
    \centering
    \includegraphics[width=1\columnwidth]{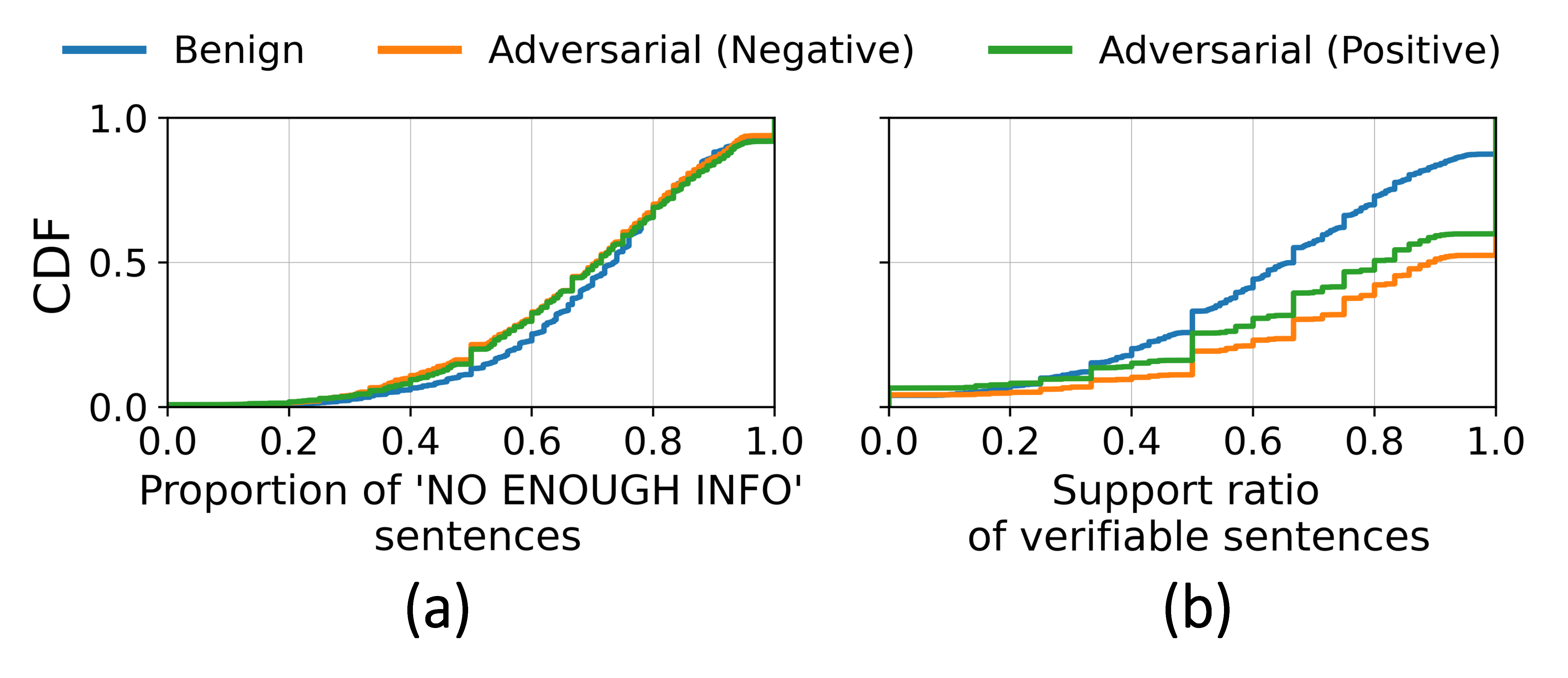}
    \captionof{figure}{\hao{Fact-checking results: (a) Adversarial passages contain comparable proportions of sentences labeled as ``NOT ENOUGH INFO'' (0.69 and 0.68 on average for positive and negative \attack\ attacks, respectively), which are similar to that of benign passages (0.71). (b) Among verifiable claims, adversarial passages contain higher proportions of sentences labeled as ``SUPPORTS'' (0.73 and 0.79 on average for positive and negative \attack\ attacks, respectively) than benign passages (0.63).}}
    \label{fig:fact}
\end{figure}
\begin{figure*}
    \centering
    \includegraphics[width=1\linewidth]{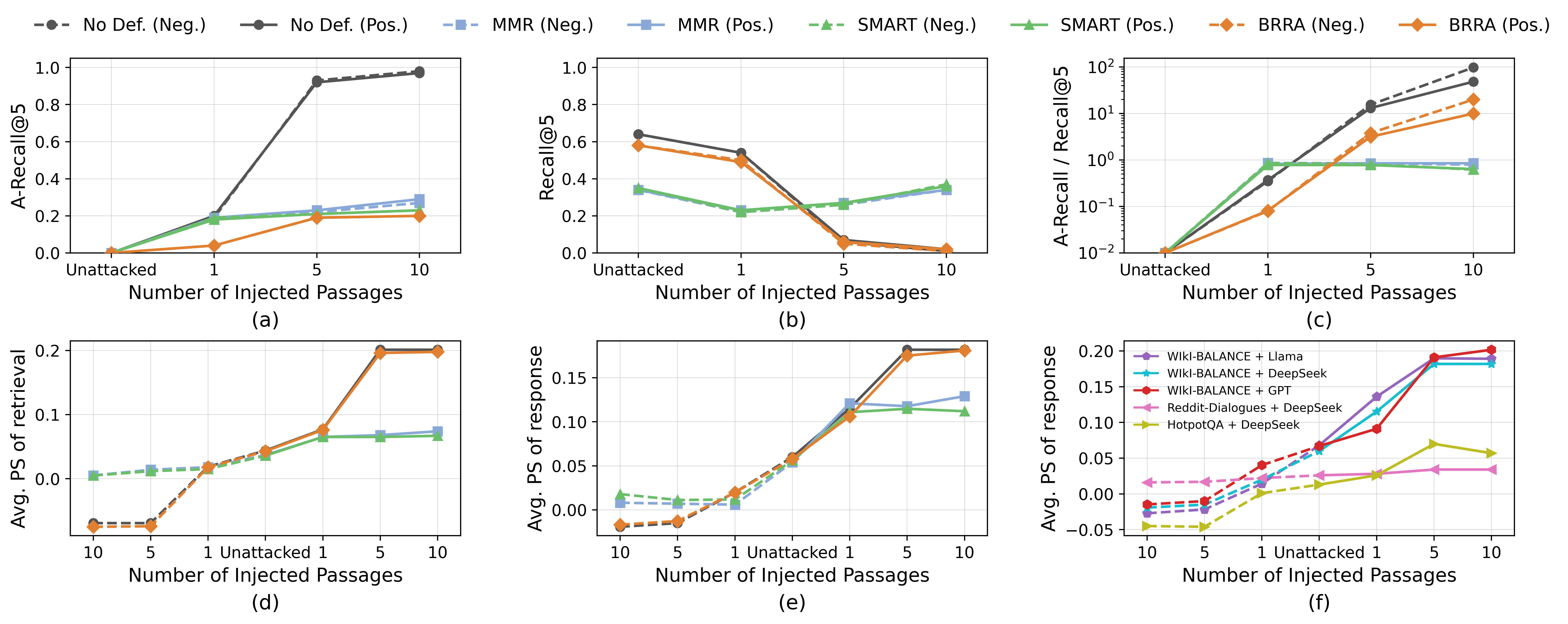}
    \caption{EBI performance against existing sanitization defenses. (a) Average A-Recall@5, the proportion of adversarial passages in top-5 retrieved passages, for varying number of attacker-injected passages. (b) Average Recall@5, the proportion of relevant benign passages (as labeled by Qrels) among the top-5 retrieved passages. (c) The ratio between average A-Recall@5 and Recall@5. (d) Average PS in top-5 retrieved passages.  (e) Average PS in LLM responses. (f) Average PS across different dataset-model combinations.}
    \label{fig:recall}
\end{figure*}
\sssec{Satisfying Property 1 (High Relevance) and Property 2 (Epistemic Bias)}. Table~\ref{tab:attackratio} shows a property check result from our benchmark, which validates the effectiveness of our attack. Specifically, our \attack\ workflow performs best on the WIKI-BALANCE dataset, with more than 86\% of the crafted passages satisfying both properties, and has descending performance on HotpotQA and Reddit-Dialogues datasets. In addition, 3.6\% to 16.0\% of the crafted passages do not have higher similarity than all the benign passages due to the existence of ultra-high-relevance benign passages in the original corpus.

Moreover, the smaller question scopes in HotpotQA and Reddit-Dialogues datasets leave limited space for epistemic bias---we cannot craft passages that have significantly different stance polarity for a nearly closed-form question---therefore causing 20\% to 37\% of crafted passages to not contain the consistent epistemic bias desired by the adversary. However, in WIKI-BALANCE with more open-ended questions, those epistemically unbiased passages are much fewer. The evaluation of epistemic bias in the responses (see Fig.~\ref{fig:recall}(f)) validates this conclusion by showing that the epistemic bias span descending ranges across those datasets. 

\sssec{Satisfying Property 3 (Stealth)}. We run the WikiCheck~\cite{WikiCheck} API on both benign and adversarial passages, which labels each sentence as ``SUPPORTS (True)'', ``REFUTES (False)'', or ``NOT ENOUGH INFO''.

As shown in Fig.~\ref{fig:fact}(a), both passage types yield a comparable proportion of ``NOT ENOUGH INFO'' labels, attributable to sentences that are non-claims (e.g., definitions, contextual or descriptive statements) or context-dependent. Thus, these sentences are unverifiable by a general-purpose knowledge base. For verifiable claims, Fig.~\ref{fig:fact}(b) shows that adversarial passages exhibit a higher support ratio (i.e., SUPPORTS / (SUPPORTS + REFUTES)) than benign ones, indicating that LLM-generated adversarial content preferentially draws on facts that fact-checkers can corroborate.

\begin{tcolorbox}[colback=green!6, colframe=black, left=1mm, right=1mm, top=1mm, bottom=1mm, boxrule=0.5pt]
\emph{Our attack creates passages that are as ``factual'' or more than benign passages to the WikiCheck fact-checker.}
\end{tcolorbox}

We note that all passages, including benign ones, contain some ``REFUTES'' labels. This reflects two limitations of per-sentence fact-checking: claims that are valid within a passage's local context may appear false when checked in isolation, and the API itself is not perfectly reliable.

\sssec{EBI vs.\ Topic-FlipRAG}. While both attacks target opinion manipulation, they serve complementary threat scenarios. Topic-FlipRAG's privileged write access and domain expertise enable precise, topic-specific manipulation---iteratively editing passages until all three requirements are met. EBI, without corpus access or domain expertise, crafts adversarial passages with no knowledge of the benign corpus, resulting in passages that do not perfectly satisfy all requirements. This is not a design shortcoming but a consequence of EBI's strictly weaker adversary assumption.

However, this imprecision makes EBI harder to defend against. Topic-FlipRAG's carefully crafted passages, by satisfying all requirements, form a cluster well-separated from benign passages in the joint space of relevance and PS, making them more amenable to detection (see Theorem~\ref{thm:optimality} in Sec.~\ref{sec:biasdef}). EBI's adversarial passages, by contrast, naturally overlap with benign ones in this space, complicating filtering without incurring excessive false positives.

\subsection{Effectiveness Against Existing Retrievers}\label{subsec:attack_ret}

We evaluate how \attack\ attacks affect different retrievers by examining two key aspects on the top-5 passages retrieved by the retrievers: A-Recall@5 and Recall@5. An effective defense should minimize A-Recall@5 as much as possible, without sacrificing the retrieval quality measured by Recall@5.

\sssec{A-Recall@5}: Fig.~\ref{fig:recall}(a) presents the average A-Recall@5 across all 452 WIKI-BALANCE queries under different injection numbers, where ``Pos.'' and ``Neg.'' denote positive \attack\ and negative \attack\ respectively. As the number of injected passages increases, all retrievers exhibit higher A-Recall@5. Among the baselines, the No Def. baseline demonstrates the highest vulnerability to \attack\ attacks, attributable to its exclusive reliance on query-passage similarity for ranking. In contrast, MMR, SMART, and BRRA incorporate perspective-based reranking that partially mitigates adversarial retrieval, yet still leave over $19\%$ of adversarial passages unsanitized at injection numbers 5 and 10. This is nearly a $5\times$ reduction compared to No Def., but it comes at the cost of reduced Recall@5 (see below).


\sssec{Recall@5}: Fig.~\ref{fig:recall}(b) presents the average Recall@5 across all 452 queries. Compared with the 0.64 of No Def. in the unattacked case, all the baselines suffer from losing 15$\sim$98\% (No Def.), 47$\sim$65\% (MMR), 42$\sim$65\% (SMART), and 22$\sim$98\% (BRRA) relevant passages and their useful content in the retrieval results across varying injection numbers. As more adversarial passages are injected (from 1 to 5 to 10), diversity-based methods (MMR and SMART) obtain larger Recall@5 values. This occurs because they penalize redundancy: When adversarial passages have similar semantic meanings, MMR and SMART avoid over-selecting from it, instead favoring more diverse passages. 


\sssec{A-Recall/Recall@5}: Since the attacker can vary the number of injected adversarial passages, we compute the ratio between A-Recall@5 and Recall@5 (Fig.~\ref{fig:recall}(c)) to assess the worst-case performance of each defense method. Specifically, the maximum ratios are as follows: 98 for No Def., 0.85 for MMR, 0.86 for SMART, and 20 for BRRA. These results demonstrate that both No Def. and BRRA are highly susceptible to adversarial passage domination, with adversarial content frequently overwhelming the top-5 retrievals. In contrast, MMR and SMART reduce this vulnerability; however, they still allow a number of unsanitized adversarial passages that is comparable to the number of relevant benign passages.

\begin{tcolorbox}[colback=green!6, colframe=black, left=1mm, right=1mm, top=1mm, bottom=1mm, boxrule=0.5pt]
\emph{Existing methods either retrieve a high proportion of adversarial passages or suppress useful benign ones in an effort to reduce adversarial retrieval. }
\end{tcolorbox}

\subsection{Impact of EBI on the LLM Responses}\label{subsec:bias}

\sssec{Calibration based on unattacked average PS}. This value, as reported in the unattacked node in Fig.~\ref{fig:recall}(d) and (e), measures the original viewpoints in unattacked cases. Specifically, the PS values of retrieved passages and responses are close to each other in unattacked cases, indicating that the LLMs usually follow the contextual stance polarity without introducing a significant model bias.

\sssec{Epistemic bias in LLM responses}. As shown in Fig.~\ref{fig:recall}(d), the attacker makes the PS of the retrieved passages biased toward its desired direction (i.e., ascending direction for positive \attack\ and descending direction for negative \attack) by injecting adversarial passages. As observed in Fig.~\ref{fig:recall}(e), such epistemic bias of retrieval can be propagated into the LLM responses. This demonstrates the vulnerability of retrieval systems to deliberate \attack\ attacks through adversarial passages. 

\sssec{Vulnerability across models and datasets}. A natural question arises: \emph{Does \attack's impact remain if we switch to a different or larger LLMs?} To answer this, we evaluate multiple LLMs using identical retrieval contexts, thereby isolating the model’s role in absorbing contextual bias. Fig.~\ref{fig:recall}(f) presents representative results for the undefended retriever on WIKI-BALANCE. Our results show that all three LLMs exhibit comparable PS shifts in their generated answers when under attack. 

The results in Fig.~\ref{fig:recall}(f) also demonstrate that our attack successfully induces PS shifts in question answering across diverse datasets. As the number of injected adversarial passages increases, the responses exhibit stronger epistemic bias, reflected by larger PS shifts across all the datasets. In addition, epistemic bias is more pronounced on WIKI-BALANCE, whose questions are more open-ended than those in HotpotQA and Reddit-Dialogues. This aligns with the observation in Table~\ref{tab:attackratio} that a higher proportion of EBI-generated passages satisfy both the relevance and epistemic bias requirements on WIKI-BALANCE.


\section{\name: A More Effective Defense}
\label{sec:biasdef}


\begin{figure}[t]
    \centering
    \includegraphics[width=\linewidth]{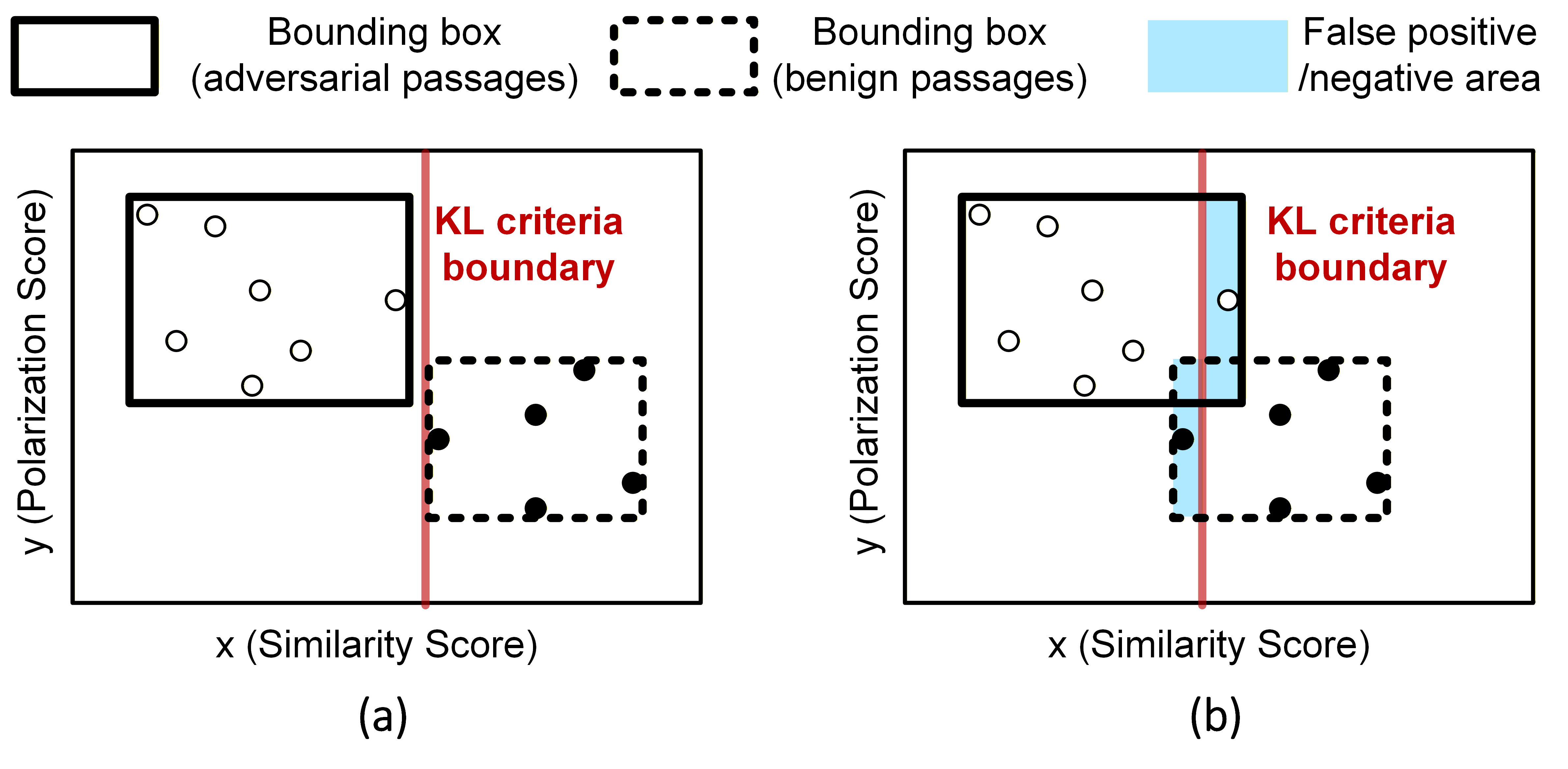}
    \caption{Two possible relative distributions of benign and adversarial passages in realistic scenarios. (a) The attacker successfully makes adversarial passages rank higher in the relevance (i.e., similarity score $\text{Sim}(q,d)$ to query). (b) Some adversarial passages are imperfectly created, losing their advantages in relevance.}
    \label{fig:defense}
\end{figure}

We have seen that existing sanitizers can attenuate the impact of \attack, but remain fundamentally insufficient. {\em Is there an adaptive defense that can defeat EBI attacks as formulated?} 

To explore this question, we design a more effective defense that explicitly leverages PS-based stance polarity quantification. The rationale behind our design is as follows: Although attackers leave no explicit surface-level fingerprints, adversarial passages exhibit statistically different distributions from benign passages when observed jointly in the two-dimensional space of similarity-to-query and PS values. Otherwise, the EBI attacks become ineffective because the Property 1 and 2 in Sec.~\ref{subsec:properties} are violated.

\subsection{Algorithmic Design}
Suppose we have a group of candidate passages in the corpus, which belong to two classes: \emph{benign} and \emph{adversarial}. We denote each passage as a node $(x,y)$ in the two-dimensional space with $x\in[0,1]$ and $y\in[-1,1]$ being the cosine similarity to the query and PS values (i.e., the projection onto the polarization axis computed by PCA on their embeddings). The $y$-distribution of nodes in the left and right of $t$ ($x<t$ and $x\ge t$) can be observed, and we denote them by $\hat{q}_L(\cdot;\,t)$ and $\hat{q}_R(\cdot;\,t))$\footnote{We use a kernel function $K(\cdot)$ to calculate the discrete kernel density of $n$ obseved nodes. For example, $\hat{q}_L(y;\,t)=\frac{1}{nh}\sum_{i=1}^nK(\frac{y-y_i}{h})$, with a uniform kernel function $K(u)=1, ||u||\le1$.}.

Then, we apply a KL criterion to find a threshold of $t$ that maximizes the Kullback–Leibler (KL) divergence~\cite{KLDivergence} between $\hat{q}_L(\cdot;\,t)$ and $\hat{q}_R(\cdot;\,t))$ as
\begin{equation}
    \hat{t}_{KL} = \arg\max_t \mathrm{KL}(\hat{q}_L(\cdot;\,t) \| \hat{q}_R(\cdot;\,t)).
\end{equation}
Then, we regard passages on the left of $\hat{t}_{KL}$ ($x< \hat{t}_{KL}$) as benign. Based on this classification criterion, we filter out the passages detected to be adversarial and returns the top-$k$ relevant passages from the rest. Algorithm~\ref{alg:DefenseDetection} describes how we implement this in real retrievers. 

\begin{algorithm}[h]
\caption{Defense Workflow of \name}
\label{alg:DefenseDetection}
\KwData{Query $q$; retrieved candidate set of passages $D$, each has $x_d$ and $y_d$; bandwidth $h_n$; small slack $\tau$ to avoid extreme boundaries}
\KwResult{Contextual passages provided to the LLM}

\For{\emph{$t$~from $\tau$ to 1-$\tau$}}{
    Compute kernel density estimators $\hat{q}_L(\cdot;\,t)$ and $\hat{q}_R(\cdot;\,t)$ for the y-distribution of nodes in the left and right of $x=t$ respectively, with bandwidth $h_n$;
}
$KL(\hat{q}_L(\cdot;\,t)||\hat{q}_R(\cdot;\,t))=\sum\limits_y \hat{q}_L(y;\,t)\log(\frac{\hat{q}_L(y;\,t)}{\hat{q}_R(y;\,t)})$;\\
$\hat{t}_{KL} = \arg\max_t \mathrm{KL}(\hat{q}_L(\cdot;\,t) \| \hat{q}_R(\cdot;\,t))$;\\
$D_{benign}\leftarrow \{d|d\in D,~x_d<\hat{t}_{KL}\}$;\\
\Return Top-$k$ passages with the highest $x_d$ in $D_{benign}$.
\end{algorithm}



\subsection{Theoretical Performance Guarantee}\label{subsec:overview}
We theoretically prove the optimality of the proposed algorithm and provide its theoretical error (i.e., false positives/negatives) boundaries. In the theorems below, \textbf{Theorem~\ref{thm:optimality}} guarantees that the KL criteria is Bayes-optimal when the attacker successfully creates adversarial passages ranking higher than benign passages in relevance to the query (see Fig.~\ref{fig:defense}(a)). \textbf{Theorem~\ref{thm:finite_sample}} gives upper bounds on the False Discovery Rate (FDR) and False Omission Rate (FOR) under estimation error in the KL criterion due to finite samples in realistic scenarios, while \textbf{Theorem~\ref{thm:overlap}} analyzes the impact of imperfectly created adversarial passages with comparable relevance to the benign passages (see Fig.~\ref{fig:defense}(b))---for example, the 7.3\% and 8.7\% passages in Table~\ref{tab:attackratio}---on the upper bounds for FDR and FOR.

\sssec{Setup and notation}. Let data points $(x_i, y_i) \in \mathbb{R}^2$ be drawn from a two-class mixture dataset. Class $k \in \{1:benign,~ 2: adversarial\}$ has prior probability $w_k$ with $w_1 + w_2 = 1$, marginal density $f_k(x)$ on $\mathbb{R}$, and conditional $y$-density $p_k(y)$. The marginal density of $x$ is $f_X(x) = w_1 f_1(x) + w_2 f_2(x)$. For any threshold $t \in \mathbb{R}$, define $F_k(t) = \int_{0}^t f_k(x)\,dx$ and $\bar{F}_k(t) = 1 - F_k(t)$. Partitioning the data at $t$ induces conditional $y$-distributions on the left ($x<t$) and right ($x\ge t$) subsets:
\begin{equation}
\begin{aligned}
  & q_L(y;\,t) = \frac{w_1 F_1(t)\,p_1(y) + w_2 F_2(t)\,p_2(y)}{w_1 F_1(t) + w_2 F_2(t)}, \\
  & q_R(y;\,t) = \frac{w_1 \bar{F}_1(t)\,p_1(y) + w_2 \bar{F}_2(t)\,p_2(y)}{w_1 \bar{F}_1(t) + w_2 \bar{F}_2(t)}.    
\end{aligned}
\end{equation}
The KL criterion is $\mathcal{L}(t) = \mathrm{KL}(q_L(\cdot;\,t) \,\|\, q_R(\cdot;\,t))$. The hard-threshold classifier at $t$ assigns $\hat{z}(x) = benign$ if $x < t$ and $\hat{z}(x) = adversarial$ if $x \ge t$. Its false positive and false negative rates are $\mathrm{FP}(t) = F_2(t)$ and $\mathrm{FN}(t) = \bar{F}_1(t)$, with corresponding posterior error rates
\begin{equation}
  \mathrm{FDR}(t) = \frac{w_2 F_2(t)}{w_1 F_1(t) + w_2 F_2(t)},~
  \mathrm{FOR}(t) = \frac{w_1 \bar{F}_1(t)}{w_1 \bar{F}_1(t) + w_2 \bar{F}_2(t)}.
\end{equation}
Throughout, $\chi^2(p_1 \| p_2) = \int p_1(y)^2/p_2(y)\,dy - 1$ denotes the chi-squared divergence between $p_1$ and $p_2$, and $\rho = \int_{0}^{1} \min(f_1(x), f_2(x))dx$ $\in [0,1]$ denotes the total-variation overlap of $f_1$ and $f_2$.
 
 
\begin{theorem}[KL Maximization is Bayes-Optimal]
\label{thm:optimality}
Suppose adversarial passages have higher similarity to a query than benign passages (\textbf{Property 1})---i.e., the two class-conditional densities on $x$ have disjoint support sets separated by $x'$:
\[
  \mathrm{supp}(f_1) \subseteq [0,\, x'), \qquad \mathrm{supp}(f_2) \subseteq [x',\, 1].
\]
Then: \emph{(a)} The criterion $\mathcal{L}(t)$ is uniquely maximized at $t = x'$, with $\mathcal{L}(x') = \mathrm{KL}(p_1 \| p_2) \ge \mathcal{L}(t)$ for all $t$, with equality if and only if $t = x'$. \emph{(b)} The hard-threshold classifier at $x'$ achieves the Bayes-optimal error: $\mathrm{FDR}(x') = \mathrm{FOR}(x') = 0$. A detailed proof is provided in Appendix~\ref{app:proofs}.
\end{theorem}


\begin{theorem}[Finite-Sample Robustness of BiasDef]
\label{thm:finite_sample}
Suppose there exists a unique KL maximizer $x'$, i.e., $\mathcal{L}(x') > \mathcal{L}(t)$ for all $t \ne x'$. Suppose $n$ i.i.d.\ pairs $(x_i, y_i)$ are observed. Let $\hat{q}_L(\cdot;\,t)$ and $\hat{q}_R(\cdot;\,t)$ be kernel density estimators of $q_L$ and $q_R$ with bandwidth $h_n = c\,n^{-1/5}$ for some constant $c > 0$, define the empirical criterion $\hat{\mathcal{L}}(t) = \mathrm{KL}(\hat{q}_L(\cdot;\,t) \| \hat{q}_R(\cdot;\,t))$, and let $\hat{t}_{KL} = \arg\max_t \hat{\mathcal{L}}(t)$. Assume: \emph{(i)} $f_1$, $f_2$ are Lipschitz at $x'$; \emph{(ii)} $f_X(x') > 0$; \emph{(iii)} $0 < \chi^2(p_1\|p_2) < \infty$. Then for every $\delta \in (0,1)$, with probability at least $1-\delta$,
\begin{equation}
  |\hat{t}_{KL} - x'| \;\le\; \frac{C\,n^{-2/5}\,\sqrt{\log(2/\delta)}}{f_X(x')\cdot\chi^2(p_1\|p_2)},  
\end{equation}
  
where $C > 0$ is a constant depending only on the kernel and Lipschitz constants. On the same event,
\begin{align}
  \mathrm{FDR}(\hat{t}_{KL}) \;\lesssim\; \frac{w_2\,f_2(x')}{w_1\,f_X(x')\cdot\chi^2(p_1\|p_2)}\cdot n^{-2/5}, \\
  \mathrm{FOR}(\hat{t}_{KL}) \;\lesssim\; \frac{w_1\,f_1(x')}{w_2\,f_X(x')\cdot\chi^2(p_1\|p_2)}\cdot n^{-2/5}.    
\end{align}
 A detailed proof is provided in Appendix~\ref{app:proofs}.
\end{theorem}
 
\begin{theorem}[FDR and FOR under $x$-Axis Overlap]
\label{thm:overlap}
Suppose $f_1$ and $f_2$ have overlap $\rho = \int \min(f_1, f_2)\,dx > 0$. Let $t^*$ be the Bayes-optimal hard threshold satisfying $w_1 f_1(t^*) = w_2 f_2(t^*)$, i.e., $\pi(t^*) = 1/2$, and let $\hat{t}_{KL} = \arg\max_t \mathcal{L}(t)$ be the population KL maximizer. Define $\delta = \hat{t}_{KL} - t^*$, $\alpha = F_1(\hat{t}_{KL})$, and $\beta = F_2(\hat{t}_{KL})$.
\begin{align}
    \mathrm{FDR}(\hat{t}_{KL}) &= \frac{w_2\,\beta}{w_1\,\alpha + w_2\,\beta}\le \frac{w_2\,\rho}{w_1\,\alpha + w_2\,\rho},\\
    \mathrm{FOR}(\hat{t}_{KL}) &= \frac{w_1\,(1-\alpha)}{w_1\,(1-\alpha) + w_2\,(1-\beta)}\le \frac{w_1\,\rho}{w_1\,\rho + w_2\,(1-\beta)}.  
  \end{align}
A detailed proof is provided in Appendix~\ref{app:proofs}.
\end{theorem}

\sssec{Summary}. The three theorems together characterize the full performance profile of the proposed detection algorithm. When the attacker successfully crafts adversarial passages with strictly higher similarity to the query than all benign passages (\textbf{Property 1}), Theorem~\ref{thm:optimality} guarantees that KL maximization is Bayes-optimal, achieving zero FDR and FOR by construction. In practice, however, two sources of error arise and manifest as increased A-Recall@5 and decreased Recall@5. First, Theorem~\ref{thm:finite_sample} shows that finite sample size introduces estimation error in the KL criterion, causing FDR and FOR that decay as $O(n^{-2/5})$ and scale inversely with $\chi^2(p_1\|p_2)$. A larger $\chi^2(p_1\|p_2)$ indicates that the adversarial passages carry stronger epistemic bias, making their PS distribution more distinguishable from that of benign passages; hence, the more aggressively an attacker biases the injected content, the more reliably our algorithm detects it. Second, Theorem~\ref{thm:overlap} shows that when some adversarial passages achieve comparable similarity to the query as benign passages---such as the 3\%--16\% of passages reported in Table~\ref{tab:attackratio}---the resulting $x$-axis overlap $\rho$ introduces an additional, irreducible component of FDR and FOR whose upper bounds grow monotonically with $\rho$.

\subsection{Effectiveness Evaluation of \name}\label{sec:biasdef-eval}
We now evaluate the efficacy of \name~against our proposed attack on the same benchmark described in Sec.~\ref{subsec:benchmark}.
\begin{figure*}
    \centering
    \includegraphics[width=\linewidth]{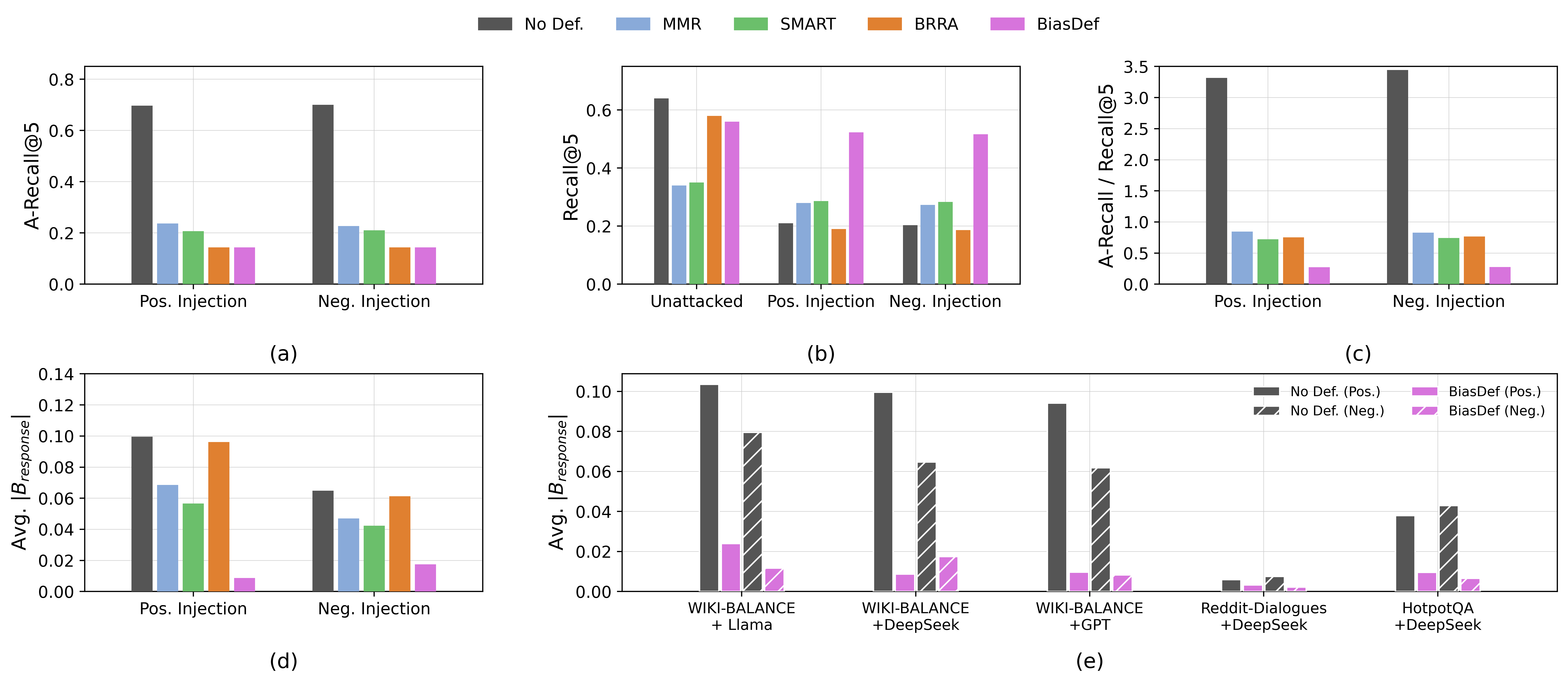}
    \caption{\name's performance. (a) Average A-Recall@5 across different injection numbers. (b) Average Recall@5 across different injection numbers. (c) The ratio of average A-Recall@5 and average Recall@5. (d) Average absolute epistemic bias $|B_\text{response}|$ in responses. (e) Average $|B_\text{response}|$ across different dataset-model combinations.}
    \label{fig:defenseResults}
\end{figure*}

\begin{figure}[t]
    \centering
    \includegraphics[width=0.85\linewidth]{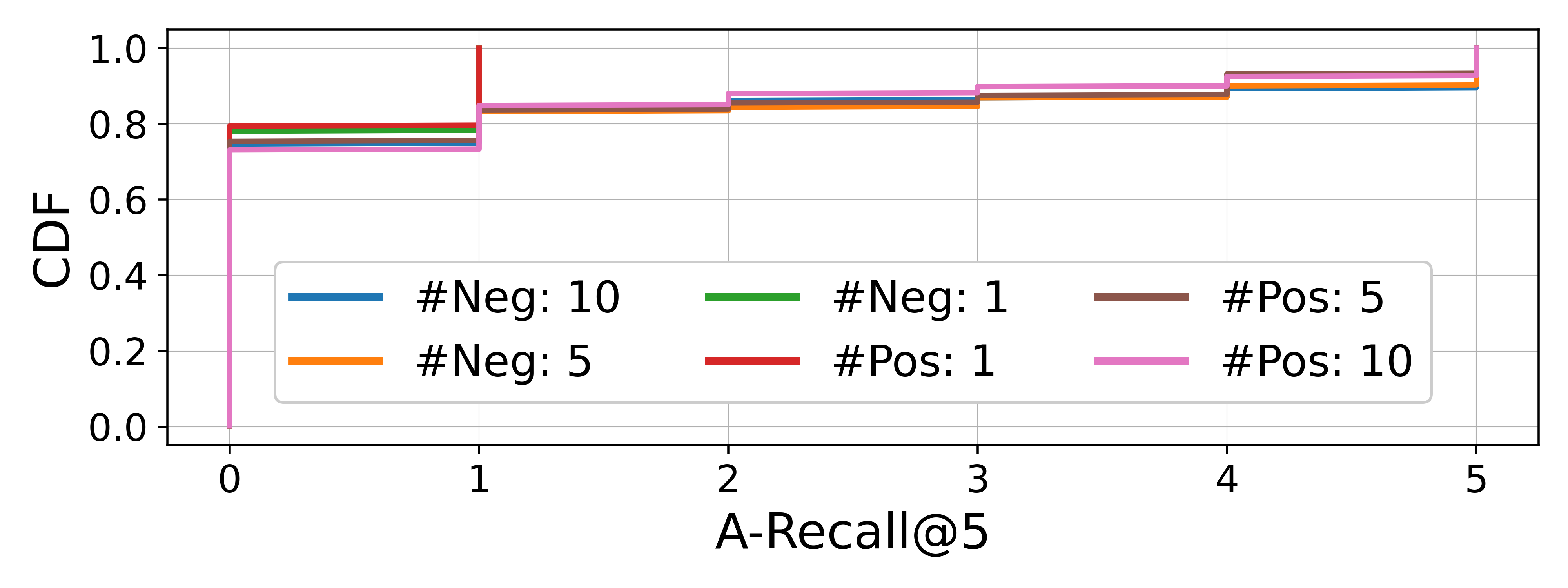}
    \caption{CDF of \name's A-Recall@5.  \name\ completely filters out adversarial passages for over 70\% queries.}
    \label{fig:CDF}
\end{figure}

\sssec{A-Recall@5}. \name\ achieves an average A-Recall@5 of 0.04, 0.19, 0.20, 0.04, 0.19, 0.20 for \#Negative Injection = 1, 5, 10, and \#Positive Injection = 1, 5, 10, respectively, which are comparable to the lowest values achieved by the baselines (BRRA in Fig.~\ref{fig:recall}(a)). Fig.~\ref{fig:defenseResults}(a) compares the averaged values across injection numbers.
A detailed analysis shown in Fig.~\ref{fig:CDF} demonstrates that \name\ is able to completely filter out adversarial passages from the top-5 results for over 70\% of the queries.

        
    
\sssec{Recall@5}. BiasDef achieves average Recall@5 values of 0.56, 0.58, 0.48, 0.49, 0.60, 0.48, and 0.49 for the unattacked case, \#Negative Injection=1,5,10, \#Positive Injection=1,5,10, respectively. Under EBI attacks, \name\ achieves the highest Recall@5 across all injection numbers among the evaluated defenses. As shown in Fig.~\ref{fig:defenseResults}(b), averaged over injection numbers, \name\ improves Recall@5 by 2.5$\times$, 1.9\%, 1.8\%, and 2.8$\times$ compared to No Def., MMR, SMART, and BRRA, respectively. Notably, a  Recall@5 reduction in the unattacked setting is a structural cost shared by all defenses.

\sssec{A-Recall/Recall@5}. \name\ reduces A-Recall@5/ Recall@5 by 12$\times$, 3.0$\times$, 2.6$\times$, 2.8$\times$ compared to No Def., MMR, SMART, and BRRA, respectively. This result demonstrates that \name\ consistently retains a low poisoning rate during retrieval. Even in the worst-case injection number of 10, the ratio of A-Recall@5 to Recall@5 is 0.4, which outperforms the worst-case performance of baselines in Sec.~\ref{subsec:attack_ret} by 2.1$\times$.


\sssec{Absolute epistemic bias in response $|B_\text{response}|$}. We observe the average $|B_\text{response}|$ in the same settings (WIKI-BALANCE + DeepSeek-R1-Distill-Qwen-14B) as in Fig.~\ref{fig:defenseResults}(d). Specifically, \name\ achieves the lowest $|B_\text{response}|$, outperforms No Def., MMR, SMART, and BRRA by 6.3$\times$, 4.3$\times$, 3.8$\times$, and 6.0$\times$, respectively. This result demonstrates that \name\ outperforms the baselines in defending EBI attacks.


\begin{haoblue}
\sssec{Generalization to LLM models and QA datasets}. Averaged over injection numbers, \name\ reduces $|B_\text{response}|$ by 5.1$\times$, 6.3$\times$, and 8.6$\times$ on Llama-3, DeepSeek-R1-Distill-Qwen, and GPT-4.1, and by 2.5$\times$ and 4.9$\times$ on Reddit-Dialogues and HotpotQA, compared to No Def. in Fig.~\ref{fig:defenseResults}(e). 

\sssec{Deployment cost}. \name\ functions as a purely plug-and-play filter, requiring no modifications to the underlying LLM. Moreover, \name\ introduces a modest additional cost (avg. of 465 \emph{ms}) for mitigating adversarial bias. This overhead is negligible compared to the typical generation time of LLMs. 
\end{haoblue}

\section{Discussion and Future Work}\label{subsec:discussion}

\begin{haoblue}

\sssec{Inherent tension in the attacker's strategy}. The
joint relevance--PS structure that EBI exploits also exposes a fundamental tension in the attacker's strategy. To maximize attack potency, adversarial passages must carry extreme PS values and high query relevance---precisely the
signatures that BiasDef detects. Conversely, if the attacker moderates the PS deviation to avoid detection, the induced
epistemic bias in the LLM response diminishes accordingly, undermining the attack's purpose. This tension is not merely
practical but structural: by Theorem~1 and Theorem~3, the
detectability of adversarial passages grows monotonically with $\chi^2(p_1 \| p_2)$, i.e., the more aggressively the attacker biases the injected content, the more reliably \name\ identifies it. Escaping this dilemma would require precise knowledge of the benign passages. Such information is unavailable without corpus access—precisely the capability our threat model denies. 

\sssec{Applicability across topic types}. The effectiveness of opinion manipulation (EBI and Topic-FlipRAG) depends on the inherent openness of the target topic. For genuinely contested questions---such as climate policy, vaccine safety, or financial regulation---a wide range of factually grounded yet epistemically divergent passages can be generated, providing ample material for effective bias injection. However, for topics with a strong factual consensus (e.g., the efficacy of well-established medical treatments or basic scientific facts), generating passages that deviate substantially from the mainstream viewpoint while remaining factually accurate becomes increasingly difficult. Our cross-dataset results in Sec.~\ref{subsec:bias} with varying topic openness reflect this. This suggests a natural boundary of applicability for EBI and Topic-FlipRAG alike, and points to topic openness as a meaningful factor in assessing the risk exposure of a given RAG deployment.

\sssec{General-purpose defense}. RAG systems operate in complex environments subject to multiple attack types, of 
which \attack\ is only one. A promising future direction is to integrate \name\ with complementary defenses targeting other threats---such as prompt injection or factual poisoning---into a unified system capable of mitigating a broader range of RAG vulnerabilities.
\end{haoblue}




\section{Conclusion}
In this paper, we show that stance polarity in passages and epistemic bias can be explicitly quantified in the embedding space, enabling a principled analysis of how retrieval outcomes shape the stance polarity of LLM-generated responses. Building on this geometric metric, we characterize \attack\ attacks on RAG systems and demonstrate their effectiveness across various retrievers, language models, and public datasets. We further present \name, a retrieval-stage defense that incorporates the same metric to filter adversarial content while largely preserving benign passages. Our results suggest that explicitly modeling stance polarity during retrieval is a promising direction for improving the robustness of RAG systems against subtle opinion manipulation.



\section*{Ethics Considerations}
Polarization and epistemic bias in LLM-generated content are growing concerns with real societal implications. Our work highlights this threat and simultaneously proposes a defense, ensuring that neither attackers nor defenders gain an asymmetrically higher advantage from our findings. All datasets used in this work are publicly available and were used solely for research evaluation purposes, with no involvement of human subjects or personally identifiable information. All experiments were conducted in a closed, local environment and had no impact on any external or production systems. We release our code and benchmark to facilitate reproducibility and to support the research community in developing stronger defenses against opinion manipulation in RAG systems.

\section*{Artifact}
We release the source code used in this paper through an anonymous GitHub repository\footnote{\url{https://github.com/OpenScience00/anonymous-EBI-artifact}}. The repository includes detailed README files that provide point-by-point instructions for constructing the benchmark and reproducing experiments.

\sssec{Environment:}
We provide all dependencies required to build the Python environment. Users only need to create a \texttt{Python~3.9.16} virtual environment and install the dependencies listed in \texttt{requirements-RAGvenv.txt}.

\sssec{Data:}
We provide the datasets used in our evaluation, including WIKI-BALANCE, HotpotQA, and Reddit-Dialogues, reorganized into an experiment-friendly format. Users who wish to evaluate our method on custom datasets may refer to the provided \texttt{synthetic\_generator.py} script in the data set directory to calculate the synthetic anchor passages necessary for the polarization axis calculation and adapt it as needed.

\sssec{Starting a local LLM server:}
To evaluate the complete RAG pipeline, users may either deploy a local LLM server or use APIs from commercial LLM providers (e.g., GPT). We provide instructions for both options. For users unfamiliar with local LLM deployment, we recommend following the Hugging Face documentation linked in the repository \texttt{README.md}.

\sssec{Running the experiments:}
To reproduce the results reported in this paper, users should execute the \texttt{run\_test.sh} script after configuring the relevant parameters. In particular, three key parameters must be specified:
\texttt{Retrieval\_Algorithm} (the defense method under evaluation),
\texttt{GeneratorType} (the LLM used for generation, which must match the model deployed on the server), and
\texttt{Dataset} (the dataset used for evaluation).
Users may optionally specify \texttt{RawDataName} to customize the filename used to store raw experimental outputs.

To process the generated raw data and obtain final evaluation metrics---including A-Recall@5, Recall@5, the mean polarization score (PS) of the top-5 retrieved passages, and the mean PS of the generated answers---users can execute the \texttt{resultprocessing.sh} script with the same \texttt{Dataset} and \texttt{RawDataName} settings.

For example, to evaluate the No Def. baseline (i.e., dense retrieval only) on the WIKI-BALANCE dataset using DeepSeek-R1-Distill-Qwen-14B as the generator model, users may use the following configuration:

\begin{verbatim}
# run_test.sh
python BiasTest.py --Mode Test \
--Retrieval_Algorithm DR \
--retrieved_num 5 \
--injection_max 10 \
--repeat_sets 1 5 10 \
--GeneratorType DeepSeek \
--RawDataName DR_results \
--Dataset WikiBalance \
--LLM_Enable 1 \
\end{verbatim}

\begin{verbatim}
# resultprocessing.sh
source RAGvenv/bin/activate
cd RAG
python ResultProcessing.py \
--Dataset WikiBalance \
--RawDataName DR_results \
\end{verbatim}

\sssec{Sources of Variability}:
During reproduction, the following factors may lead to discrepancies with the results reported in this paper.

First, the effectiveness of both the baseline defenses and our proposed method depends on the quality of adversarial passages generated by the attacker. When users regenerate adversarial passages using their own LLMs, the resulting passages may differ from those provided by the authors due to the inherent stochasticity of LLM decoding (e.g., sampling randomness). Consequently, identical numerical results (e.g., A-Recall@5) are not guaranteed. To minimize this variance, we recommend reproducing the reported defense performance using the adversarial passages provided in our repository, while using custom LLMs primarily to validate the end-to-end attack workflow.

Second, although the retrieval results can be deterministically reproduced, the PS of generated answers may vary across different execution environments. Such variations can arise from differences in LLM randomness as well as hardware and system configurations, including GPU type, GPU count, and parallelism settings. In this case, users are expected to reproduce the \emph{relative trends} of answer PS—rather than exact numerical values—across different injection intensities, datasets, and defense methods, which is sufficient to validate the conclusions of this paper.

We have not developed or trained any LLM. All use of LLMs is in compliance with intellectual property rights or terms of public use. Our dataset is Wikipedia and Reddit. We have only tested properties of publicly available models / API on public datasets.

\bibliographystyle{IEEEtran}
\bibliography{reference.bib}

\appendices
\section{Prompt Used}\label{app:prompt}
An example prompt used in typical RAG process is as follows.




\begin{tcolorbox}[colback=gray!10!white, colframe=black, left=1mm, right=1mm, top=1mm, bottom=1mm, boxrule=0.5pt]
\noindent
\texttt{You are a helpful assistant. Use the following context to answer the question as accurately and objectively as possible.}

\vspace{0.1in}
\texttt{Context:
[1] \{Passage 1\}; …  [k] \{Passage k\}}

\vspace{0.1in}
\texttt{Question: \{query\}}

\vspace{0.1in}
\texttt{Answer: }
\end{tcolorbox}

The prompts used to generate synthetic passages, similar passages to the mostly biased synthetic passage and the final answers are as follows.

\sssec{Prompt-Synthetic}:
\begin{tcolorbox}[colback=gray!10!white, colframe=black, left=1mm, right=1mm, top=1mm, bottom=1mm, boxrule=0.5pt]
    \noindent
    \texttt{Instruction: Write \{J\} persuasive and diverse articles answering the user question below.
    Each article should present a **different viewpoint or ideological stance**, covering a range of perspectives.
    The responses should sound natural and plausible, as if written by different authors with distinct opinions.}

    \vspace{0.1in}
    \texttt{User Question: \{Query\}}

    \vspace{0.1in}
    \texttt{Output: }
\end{tcolorbox}

\sssec{Prompt-Adv}:
\begin{tcolorbox}[colback=gray!10!white, colframe=black, left=1mm, right=1mm, top=1mm, bottom=1mm, boxrule=0.5pt]
    \noindent
    \texttt{Instruction: Write a persuasive article answering the user question below.
    Your article must follow the same stance and bias intensity as the example passage.
    Avoid copying text, but ensure the tone, viewpoint, and strength are closely aligned.}
    
    \vspace{0.1in}
    \texttt{User Question: \{Query\}}
    
    \vspace{0.1in}
    \texttt{Example Passage: \{$d_\text{bias}$\}}
    
    \vspace{0.1in}
    \texttt{Output: }
    \end{tcolorbox}

\sssec{Prompt-Generation}:
\begin{tcolorbox}[colback=gray!10!white, colframe=black, left=1mm, right=1mm, top=1mm, bottom=1mm, boxrule=0.5pt]
    \noindent
    \texttt{Instruction: You are an expert who has just read several passages about a controversial issue. Your job is to read these passages, carefully consider their arguments and evidence, and then write a clear, well-justified answer to the following question.
    **Do not simply summarize or list the passages. Instead, think critically and use the information to form your own conclusion.**}
    
    \vspace{0.1in}
    \texttt{User Question: \{Query\}}
    
    \vspace{0.1in}
    \texttt{Contextual Passages: [1] \{Passage 1\}; …  [k] \{Passage k\}}
    
    \vspace{0.1in}
    \texttt{Answer: }
    \end{tcolorbox}

\section{Existing Sanitizers and Their Weaknesses}\label{subsec:existingdefense}

The most basic countermeasures against \attack\ to consider are those which rely solely on maximizing the relevance (i.e., simlarity to the query) of the retrieved passages. Along this line of defenses, dense-retrieval methods such as DPR~\cite{DR} and Sentence-BERT~\cite{SBERT}, are widely adopted in RAG systems due to their effectiveness in capturing semantic similarity. However, they typically do not have any built-in defense mechanisms against \attack. Because retrieval decisions are driven solely by  relevance, adversarial passages that are both biased and highly similar to the query can be easily retrieved. Our undefended baseline employs these mechanisms by default.

\begin{figure}[t]
    \centering
    \includegraphics[width=0.98\linewidth]{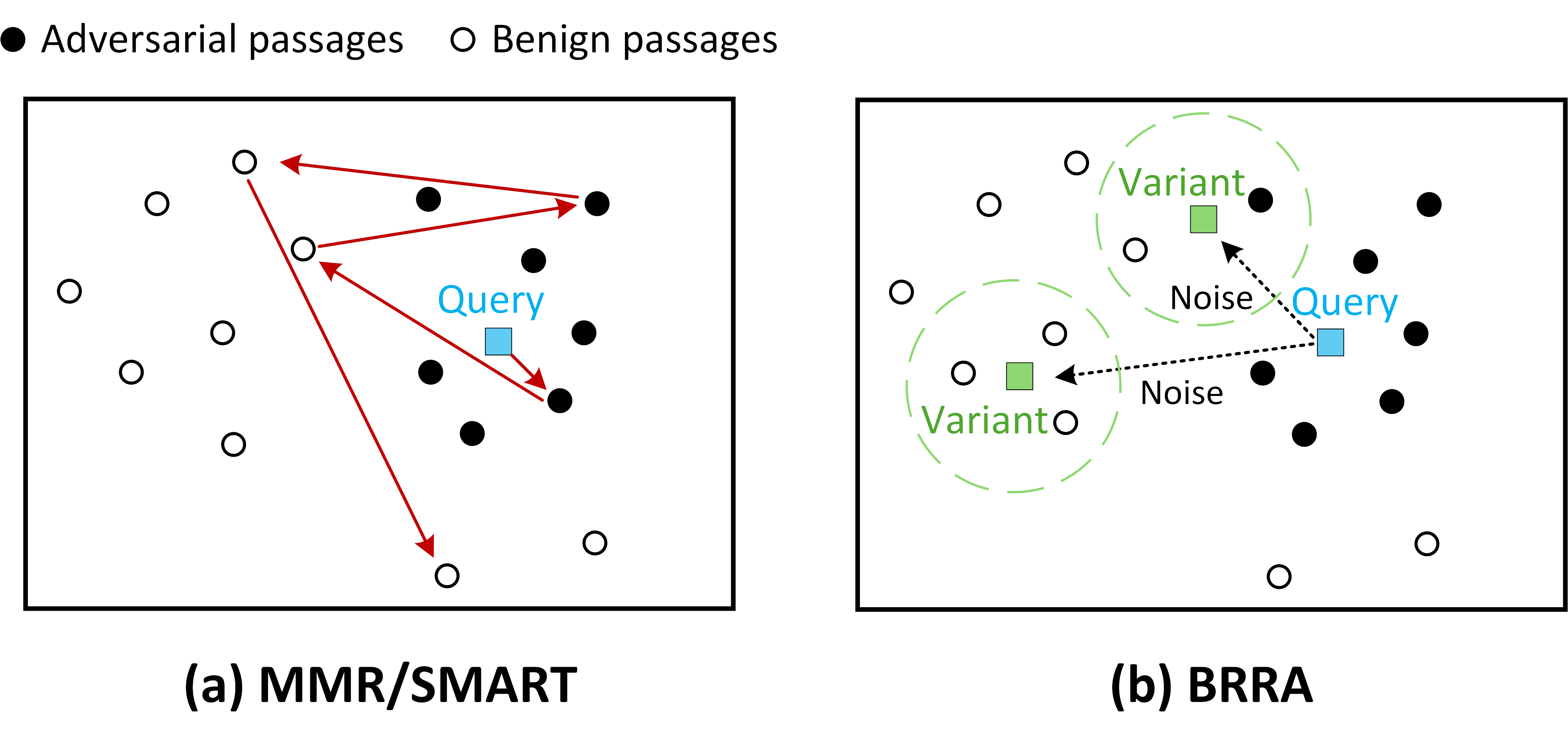}
    \caption{Illustration of state-of-the-art retrieval strategies in the embedding space. 
    (a) As indicated by the red arrows, MMR~\cite{MMR} and SMART~\cite{SMART-RAG} tend to favor diversity by selecting passages that are not necessarily the most similar to the query. (b) BRRA~\cite{biasamplify} retrieves passages relevant to both the original query and its noise-perturbed variants, then re-ranks them. As a result, benign passages near the perturbed queries (shown as green circles) may be retrieved even if they are distant from the original query.
}
    \label{fig:existing}
\end{figure}

In contrast to these relevance-only retrievers, several perspective-aware retrieval methods have been proposed to balance relevance, diversity, and conflict among the retrieved passages. They can serve as potential defense mechanisms for \attack\ attacks. 
Prominent examples of these include MMR~\cite{MMR} and SMART~\cite{SMART-RAG}.
MMR reduces redundancy by penalizing candidates that are highly similar to already selected passages. It iteratively selects passages based on a relevance-diversity score, which accounts for both similarity to the query and dissimilarity to previously selected passages, thereby encouraging content diversity. SMART employs a more sophisticated selection strategy to optimize the trade-off among relevance, inter-passage similarity, and conflict resolution. Although these methods are effective in avoiding repetition and limiting the dominance of adversarial passages, as we will show later, they cannot fully eliminate EBI adversarial content---the diversity mechanism inadvertently encourages the retrieval of both benign and adversarial passages  (see Fig.~\ref{fig:existing}(a)).


Lastly, unlike the above methods, BRRA~\cite{biasamplify} introduces query perturbation to enhance diversity. It generates multiple noisy perturbations to the query and retrieves passages relevant to both the original and perturbed queries (illustrated as green circles in Fig.~\ref{fig:existing}(b)). It then re-ranks all retrieved passages based on retrieval frequency and rank to promote coverage across the semantic space. BRRA is effective in low-dimensional embedding spaces, where noise-induced perturbations cause significant angular deviation. However, in high-dimensional settings, random noise induces only minor directional shifts after normalization, due to the concentration of measure on the unit hypersphere~\cite{Vershynin_2018}. As a result, the similarity between original and perturbed queries remains high (e.g., $>$0.9), yielding minimal variation in retrieved results.

\section{Proof of Theorems}\label{app:proofs}
\sssec{Proof of Theorem~\ref{thm:optimality}} 
\begin{proof}
\textbf{(a)} Since $\mathrm{supp}(f_1) \subset [0, x')$ and $\mathrm{supp}(f_2) \subset [x', 1]$, we have $F_1(x') = 1$, $F_2(x') = 0$, and hence $q_L(y;\,x') = p_1(y)$, $q_R(y;\,x') = p_2(y)$, giving $\mathcal{L}(x') = \mathrm{KL}(p_1 \| p_2)$. For $t < x'$: the support conditions imply $F_2(t) = 0$, so $q_L(y;\,t) = p_1(y)$, while the right subset mixes $benign$ mass from $[t, x')$ and all $adversarial$ mass, giving
\begin{align*}
q_R(y;\,t) &= \alpha(t)\,p_1(y) + (1-\alpha(t))\,p_2(y), \\
\alpha(t) &= \frac{w_1(1-F_1(t))}{w_1(1-F_1(t))+w_2} \in (0,1).
\end{align*}
The function $\phi(\alpha) = \mathrm{KL}(p_1 \| \alpha p_1 + (1-\alpha) p_2)$ is convex in $\alpha \in [0,1]$ since $\mathrm{KL}(p\|\cdot)$ is convex, with boundary values $\phi(0) = \mathrm{KL}(p_1\|p_2)$ and $\phi(1) = 0$. By convexity, $\phi(\alpha(t)) \le (1-\alpha(t))\,\mathrm{KL}(p_1\|p_2) < \mathrm{KL}(p_1\|p_2)$ for all $\alpha(t) \in (0,1)$, so $\mathcal{L}(t) < \mathcal{L}(x')$. 

For $t > x'$: $\bar{F}_1(t) = 0$ gives $q_R(y;\,t) = p_2(y)$, and the left subset satisfies $q_L(y;\,t) = \beta(t)\,p_1(y) + (1-\beta(t))\,p_2(y)$ for some $\beta(t) \in (0,1]$. Convexity of $\mathrm{KL}(\cdot \| p_2)$ yields $\mathrm{KL}(q_L(t)\|p_2) \le \beta(t)\,\mathrm{KL}(p_1\|p_2) < \mathrm{KL}(p_1\|p_2)$. This establishes strict inequality for all $t \ne x'$.

\textbf{(b)} The disjoint-support assumption gives $P(Z=benign\mid x)=1$ for $x \in \mathrm{supp}(f_1)$ and $P(Z=adversarial\mid x)=1$ for $x \in \mathrm{supp}(f_2)$. The classifier at $x'$ achieves $F_2(x')=0$ and $\bar{F}_1(x')=0$, hence $\mathrm{FDR}(x')=\mathrm{FOR}(x')=0$.
\end{proof}
 

\sssec{Proof of Theorem~\ref{thm:finite_sample}} 
\begin{proof}
\textbf{Step 1: Uniform KDE error.} For a fixed threshold $t$, the left and right subsets contain $n_L \sim \mathrm{Bin}(n, F_X(t))$ and $n_R = n - n_L$ samples. By the standard KDE bias-variance decomposition, the $L^2$ error of $\hat{q}_k$ satisfies $\mathbb{E}[\|\hat{q}_k - q_k\|_2^2] = O(h_n^4) + O(1/(n_k h_n))$, balanced at $h_n = c\,n^{-1/5}$ to give $O(n^{-4/5})$. Since $\chi^2(p_1\|p_2) < \infty$, the densities $q_L$ and $q_R$ are mutually absolutely continuous, so $\min_y q_R(y;\,t)$ is bounded away from zero near $x'$. The induced error in the KL estimate is therefore
\[
  |\hat{\mathcal{L}}(t) - \mathcal{L}(t)| = O\!\left(\frac{\|\hat{q}_L - q_L\|_2 + \|\hat{q}_R - q_R\|_2}{\min_y q_R(y;\,t)}\right) = O(n^{-2/5})
\]
uniformly over the $O(n)$ candidate thresholds near $x'$, by a union bound over a covering of the search interval.

\textbf{Step 2: KL gradient at $x'$.} For $t = x' - \varepsilon$, $\varepsilon > 0$ small, the right-subset mixing weight is $\alpha(t) = w_1(1-F_1(t))/(w_1(1-F_1(t))+w_2\bar{F}_2(t))$. Since $f_X(x') > 0$, both $f_1$ and $f_2$ contribute mass near $x'$, and $\alpha(t) \approx w_1 f_1(x')\varepsilon / f_X(x')$ to first order. Differentiating $\phi(\alpha) = \mathrm{KL}(p_1 \| \alpha p_1 + (1-\alpha)p_2)$ at $\alpha = 0$ gives $\phi'(0) = -\chi^2(p_1\|p_2)$, so
\[
  \mathcal{L}(x') - \mathcal{L}(x'-\varepsilon) \;\approx\; \varepsilon \cdot \frac{w_1 f_1(x')}{f_X(x')}\cdot\chi^2(p_1\|p_2).
\]
Since $f_X(x') > 0$ and $\chi^2(p_1\|p_2) > 0$, the local gradient magnitude is $\kappa = f_X(x')\cdot\chi^2(p_1\|p_2)\cdot\min(w_1,w_2) > 0$.

\textbf{Step 3: Argmax localization.} By the argmax stability lemma, if the population criterion has gradient $\kappa > 0$ at $x'$ and the supremum estimation noise is $\sigma_n = O(n^{-2/5})$, then with probability $1-\delta$,
\[
  |\hat{t}_{KL} - x'| \;\le\; \frac{\sigma_n}{\kappa}\sqrt{\log(2/\delta)} = \frac{C\,n^{-2/5}\sqrt{\log(2/\delta)}}{f_X(x')\cdot\chi^2(p_1\|p_2)}.
\]

\textbf{Step 4: FDR and FOR.} Write $\hat{t}_{KL} = x' + \delta_n$. By Lipschitz continuity of $f_1$ and $f_2$ at $x'$, $F_2(x'+\delta_n) - F_2(x') \le f_2(x')|\delta_n|$ and $F_1(x') - F_1(x'+\delta_n) \le f_1(x')|\delta_n|$. Substituting into the exact formulas $\mathrm{FDR}(\hat{t}) = w_2 F_2(\hat{t})/(w_1 F_1(\hat{t})+w_2 F_2(\hat{t}))$ and $\mathrm{FOR}(\hat{t}) = w_1\bar{F}_1(\hat{t})/(w_1\bar{F}_1(\hat{t})+w_2\bar{F}_2(\hat{t}))$, and using the localization bound from Step 3, yields the stated FDR and FOR rates. 
\end{proof}
 
\begin{remark}
Assumption \emph{(ii)} ($f_X(x') > 0$) requires that data from both classes be present near $x'$, so the KL criterion can detect the distributional change. This is automatically satisfied when $f_1$ and $f_2$ overlap near $x'$ (as in Theorem~\ref{thm:overlap}), but is violated under the strict disjoint-support assumption of Theorem~\ref{thm:optimality}: if $x'$ lies in a vacuum band, the algorithm achieves exact zero error trivially and the finite-sample rates of this theorem are unnecessary; if $\chi^2(p_1\|p_2) = +\infty$, the conclusion still holds at a rate no slower than $O(n^{-2/5})$ by a direct argument. Theorem~\ref{thm:finite_sample} therefore describes the non-trivial regime where both the signal ($\chi^2$) and the local data density ($f_X(x')$) are finite and positive.
\end{remark}

\sssec{Proof of Theorem~\ref{thm:overlap}}
\begin{proof}
\textbf{Exact formulas.} By Bayes' theorem, the FDR is the posterior probability that a point predicted as $benign$ (i.e., $x < \hat{t}_{KL}$) actually belongs to $adversarial$:
\begin{align*}
  &\mathrm{FDR}(\hat{t}_{KL}) = P(Z=adversarial\mid\hat{z}=benign) = \\
  &\frac{P(\hat{z}=benign\mid Z=adv.)\,w_2}{P(\hat{z}=benign\mid Z=benign)\,w_1 + P(\hat{z}=benign\mid Z=adv.)\,w_2} \\
  &= \frac{w_2 F_2(\hat{t}_{KL})}{w_1 F_1(\hat{t}_{KL})+w_2 F_2(\hat{t}_{KL})}.
\end{align*}
The FOR formula follows identically with $F_k$ replaced by $\bar{F}_k$ and the roles of the two classes swapped.

\textbf{Upper bounds.} Only mass in the overlap region contributes false positives, so $F_2(\hat{t}_{KL}) = \int_{-\infty}^{\hat{t}_{KL}} f_2(x)\,dx \le \int_{-\infty}^{\infty} \min(f_1(x), f_2(x))\,dx = \rho$. Substituting $F_2(\hat{t}_{KL}) \le \rho$ into the exact FDR formula, and using $w_1\alpha \ge 0$:
\begin{align*}
  \mathrm{FDR}(\hat{t}_{KL}) = \frac{w_2 F_2(\hat{t}_{KL})}{w_1\alpha + w_2 F_2(\hat{t}_{KL})} \le \frac{w_2\,\rho}{w_1\,\alpha + w_2\,\rho}.  
\end{align*}
Symmetrically, $\bar{F}_1(\hat{t}_{KL}) = \int_{\hat{t}_{KL}}^{\infty} f_1(x)\,dx \le \rho$, giving
\[
  \mathrm{FOR}(\hat{t}_{KL}) = \frac{w_1\bar{F}_1(\hat{t}_{KL})}{w_1\bar{F}_1(\hat{t}_{KL})+w_2(1-\beta)} \le \frac{w_1\,\rho}{w_1\,\rho + w_2\,(1-\beta)}. 
\]
\end{proof}

\end{document}